\documentclass[a4paper,12pt]{article}

\title {\LARGE\bf   Random incidence matrices:
\\                  moments of the spectral density}

\author {           M. Bauer and O. Golinelli
\bigskip
\\ \ad              Service de Physique Th\'eorique, Cea Saclay 
\\ \ad              91191 Gif-sur-Yvette, France
\\ \ad              email: bauer and golinelli@spht.saclay.cea.fr
}
\date{\normalsize   July 6, 2000
\\                  revised: October 13, 2000
\\\mbox{}\\         Preprint T00/088 ; cond-mat/0007127
}

\newcommand  {\ad}{\normalsize\em}      % style for address

\usepackage[final]{graphicx}
\newcommand{\figwidth}{11.3truecm}       % width of figures

\newcommand{\Tr}{{\rm Tr}}
\newcommand{\tr}{{\rm tr}}
\newcommand{\CD}{{\cal D}}
\newcommand{\CH}{{\cal H}}
\newcommand{\CI}{{\cal I}}
\newcommand{\CL}{{\cal L}}
\newcommand{\CS}{{\cal S}}
\newcommand{\al}{\alpha}
\newcommand{\om}{\omega}
\newcommand{\Om}{\Omega}
\newcommand{\la}{\lambda}
\newcommand{\ket}[1]{{|#1\rangle}}
\newcommand{\bra}[1]{{\langle#1|}}
\newcommand{\bin}[2]{{\left(\!\!\begin{array}{c}#1\\#2\end{array}\!\!\right)}}

\begin{document}

\maketitle

\begin{abstract}
%================

We study numerically and analytically the spectrum of incidence matrices of
random labeled graphs on $N$ vertices : any pair of vertices is connected
by an edge with probability $p$.  We give two algorithms to compute the
moments of the eigenvalue distribution as explicit polynomials in $N$ and
$p$.  For large $N$ and fixed $p$ the spectrum contains a large eigenvalue
at $Np$ and a semicircle of ``small'' eigenvalues.  For large $N$ and
fixed average connectivity $pN$ (dilute or sparse random matrices limit) we
show that the spectrum always contains a discrete component. An anomaly in
the spectrum near eigenvalue $0$ for connectivity close to $e$ is
observed. We develop recursion relations to compute the moments as explicit
polynomials in $pN$. Their growth is slow enough so that they determine the
spectrum. The extension of our methods to the Laplacian matrix is given in
Appendix.


\end{abstract}

\section{Introduction}
%=====================

The spectral properties of the incidence matrix of random graphs have
motivated a large number of studies over the last decades. The same problem
is described under rather different names, depending upon the aspects that
are under focus and the method of attack.

The interest in this problem has several roots in physics. The replacement
of complicated hamiltonians by large random matrices has proved very
efficient in the analysis of the spectral properties (culminating in level
spacing distributions) of large nuclei since the pioneering works of Wigner
and Dyson. For many properties, the details of the probabilistic laws
governing the distribution of matrix elements are irrelevant, and there is
a very powerful notion of universality. Further motivation for considering
precisely random incidence matrices comes from several systems in condensed
matter physics, a good example being conductors with impurities. The pure
system is modeled by a lattice, and electrons can move along
bonds. Impurities break bonds. So the hamiltonian can be approximated by
the incidence matrix of the lattice with random bonds removed. If one
considers several large samples differing by the impurity concentration
$c$, the following properties are observed. When $c$ is large, only small
islands of metallic atoms exist. If $c$ decreases to reach a certain
threshold, a large island of metallic atoms invades the system. This is
classical percolation. However it is generally believed that the system
remains insulating (the wave function of the electrons are all localized)
until another threshold in the impurity concentration. Then some
delocalized states appear and the sample is a conductor. This is called
quantum percolation, a kind of Anderson (de)localization. The analytic
study of this problem on a 3d lattice with random bonds removed is very
difficult, and this motivated people to look at the much simpler problem of
a random graph. This forgets about the spatial structure and is a kind of
mean field approximation.

In the random graph model, the lattice is replaced by the complete graph on
$N$ points: any two points are connected by a bond (by an edge in the
language of graph theory). Then, bonds are randomly removed, leading to a
random graph where only a fraction $p$ of the initial bonds remains.  In
the simplest case, bonds are removed with probability $1-p$ independently
of each other. The model can be made more complicated by choosing randomly
a sign for each bond present in the random graph. This allows interferences
if the probability amplitude for an electron moving on the random graph is
the product of the signs of the visited bonds.

The topology of large random graphs was investigated about four decades ago
by Erd\"os and Renyi~\cite{erdos} in a remarkable series of papers. The
idea is to let $p$ vary with $N$. There are quite a few different
regimes. The most relevant for further physical investigations are:
\begin{itemize}
  \item The edge-probability $p$ remains fixed as $N$ goes to infinity.

  \item The average connectivity $\alpha=pN$ remains fixed as $N$ goes to
infinity. 
\end{itemize}

In the first case, the infinite random graph is connected and in a precise
sense two infinite random graphs of given $p$ are isomorphic with
probability 1. The second case exhibits a percolation transition at $\alpha
=1$. For small $\alpha$ all connected components are finite, and only trees
contribute to the extensive quantities. But for $\alpha >1$, a finite
fraction of the points lies in a single connected component.

As explained above, the spectral properties of the (signed) incidence
matrix of the random graph, a symmetric matrix with $0,(\pm)1$ matrix
elements have a great physical interest. Quite often, authors concentrate
on the case when the distribution of random signs is symmetric. Numerical
simulations and analytic (mostly supersymmetric) methods have given a great
amount of information.

For large $N$, fixed $p$ and symmetric random signs, it is known that the
spectral distribution is a semicircle~\cite{rodgers88} and that the level
correlations are those of the Gaussian orthogonal ensemble (GOE), one of
the four standard universality classes governing random
spectra~\cite{mirlin91}.

The finite connectivity limit has also attracted a lot of attention, under
the names of dilute or sparse random matrices. It has been
argued~\cite{mirlin91,evangelou92,evangelou92a} that there is some value
$\alpha_q>1$ for which delocalized eigenstates appear. So this simple model
is believed to exhibit a quantum percolation transition.

Our aim in this paper is to use combinatorial methods to compute explicitly
moments of the spectral distribution for given $N$ and $p$, and in the
finite connectivity limit.

Section 2 gives a formal definition of the model and recalls some of its
topological properties. Section 3 gives the enumerative algorithm for
moments as polynomials in $N$ and $p$. Section 4 concentrates on the fixed
$p$ large $N$ limit, first numerically (spectrum and level spacing) and
then analytically. Section 5 deals with the finite connectivity large $N$
limit, starting with numerical computations. In particular, we observe a
quantitative change in the spectrum near the eigenvalue $\lambda=0$ for
$\alpha \simeq 2.7$. We give qualitative arguments for the presence,
location and size of delta peaks in the spectrum. Then we derive a formal
expression for the moment generating function and give a recursion relation
for the moments, that we use to control their growth. In an Appendix, we
show how our algorithms can be extended when the incidence matrix is
replaced by the Laplacian matrix of a random graph.

\section{The model}
%==================

\label{sec:model}

For $N=1,2,\cdots$, we define the sample space $\Omega_N$ as follows : the
elements of $\Omega_N$ are the symmetric $N \times N$ matrices
$M=(M_{ij})_{i,j\in[1,N]}$, with $M_{i,i}=0$ on the diagonal and
$M_{i,j}=0$ or 1 for $i\neq j$.  So $\Omega_N$ is a discrete space
consisting of $2^{N(N-1)/2}$ points. We observe that $\Omega_N$ is in one
to one correspondence with the set of labeled \textit{simple} graphs on $N$
vertices : to $M \in \Omega_N$, we associate the graph with vertex set
$\{1,\cdots N\}$ and edge set $E(M)=\{\{i,j\} \mid M_{ij}=1\}$.  Then
$|E(M)| =\frac{1}{2}\sum_{i,j} M_{ij}=\sum_{i<j} M_{ij}$ is just the number
of edges of the graph associated to $M$. The word \textit{simple} above
refers to the fact that the graphs we consider do not contain multiple
edges or edges with only one vertex. In the sequel, graph always means
simple graph, and we talk indiscriminately of matrices or associated
graphs.

For any $p \in [0,1]$ we turn $\Omega_N$ into a probability space : the
weight of $M \in \Omega_N$ is $P(M)=p^{E(M)}(1-p)^{N(N-1)/2-E(M)}$. To
rephrase this formal definition, the entries of $M$ above the main diagonal
are independent random variables with the same Bernoulli (binomial)
distribution: for $i < j$, $M_{ij} = 1$ (or equivalently the vertices $i$
and $j$ are connected by an edge) with probability $p$ and $M_{ij} = 0$
with probability $1-p$.

The quantity $\alpha \equiv pN$ appears as the average
connectivity\footnote{The terminology \textit{average connectivity} seems
to be well established in the physics literature, and we stick to it. It
would however be more consistent with general graph theory to call it the
\textit{average degree} (of vertices)},
i.e. the average number of vertices $j$ connected by an edge $\{i,j\}$ to a
given vertex $i$.  Remark that this connectivity fluctuates, in contrast
to regular graphs~\cite{jakobson} where the connectivity is fixed for all
vertices.

\label{sec:variant}
We can define a variant of this model by introducing a random sign, with a
parameter $a\in [0,1]$: for $i<j$, $M_{ij} = +1$ with probability $ap$,
$M_{ij} = -1$ with probability $(1-a)p$ and $M_{ij} = 0$ with probability
$1-p$.  The even model $a = 1/2$, which gives $\langle M_{ij}\rangle = 0$,
has been studied by some
authors~\cite{rodgers88,rodgers90,mirlin91,evangelou92,evangelou92a}.  If a
random graph contains no loops (i.e. closed circuits), the parameter $a$ is
not relevant because all the negative signs can be changed in positive ones
by a simple change of basis.  More generally, it is true if the graph has
no ``frustrated'' loops, i.e. no loops with odd number of negative edges.
We will see later that the random spectrum is not sensitive to $a$ in the
large $N$ limit with fixed $\alpha$.  So, without explicit indications, we
will speak about the signless model defined previously, which has $a=1$.

If $X$ is any random variable on $\Omega_N$, we use the notation
$\overline{X}$ for the expectation value (or average) of $X$.  In the next
sections, we shall be interested in the asymptotic behavior of the moments
of the spectral density $\overline{\Tr \; M^k}$ when $N \rightarrow
\infty$, first with $p$ fixed and then with $\alpha =Np$ fixed. But first,
we recall a few fundamental facts on the topology of random graphs.  The
basic reference is \cite{erdos}; for a textbook presentation and more
references, see e.g. \cite{bollobas}.

\label{sec:perco}
It is well established that this model has a {\em percolation} transition at
$\alpha = 1$.  In the regime $\alpha < 1$, with probability 1 in the large
$N$ limit, all the connected components of the random graph are finite:
moreover they are mostly trees, there is only a finite number of one loop
components and no other connected components.  For a given tree $T$ on $n$
vertices, the average number of connected components isomorphic to $T$ is
\begin{equation}
  \frac{N}{|Aut(T)|}\alpha^{n-1} e^{-n\alpha} + o(N)
  \label{eq:probtree}
\end{equation}
where $|Aut(G)|$ for a given graph $G$ is defined as the order of its
automorphism group, \label{sec:aut} formed by the permutations of the
vertices that leave its incidence matrix invariant.

In the regime $\alpha > 1$, Eq~(\ref{eq:probtree}) remains valid, but one
``giant'' connected component (equivalent of the percolation cluster for
regular lattices) appears, with a finite fraction of the $N$ vertices and
many loops.  This fraction is an increasing function of $\alpha$, covering
$[0,1]$ when $\alpha$ runs from 1 to $\infty$.  In the limit $N$ large with
$p$ fixed, the random graphs is made only with one component.

The percolation transition at $\alpha=1$ is of second order.  As usual,
critical exponents can be defined: by example, the biggest component has a
size of order $N^{2/3}$.

Moreover this model exhibits~\cite{mirlin91} an Anderson {\em localization}
transition, also called {\em quantum percolation} transition, at a value
$\alpha_q>1$.  In the phase $\alpha<\alpha_q$, all eigenvectors of the
random incidence matrix are localized.  On the other hand, for
$\alpha>\alpha_q$, the eigenvectors for which the absolute value of the
energy is below a threshold $E(\alpha)$ becomes extended.

By studying nearest level spacing between eigenvalues, we expect an
exponential distribution of spacing in the localized phase (because the
spatial covering between different eigenvectors vanishes), and a GOE
distribution~\cite{mehta} in the delocalized phase.  With this kind of
criterion, the location transition has been numerically
estimated~\cite{evangelou92,evangelou92} at $\alpha_q
\approx 1.4$.  By considering quantum percolation on a randomly diluted 
Cayley (or Bethe) tree~\cite{evangelou92,evangelou92a,harris}, it has been
conjectured that $\alpha_q$ is given by $\alpha_q \log \alpha_q = 1/2$,
(leading to $\alpha_q \approx 1.4215299$).  We argued~\cite{bauer00a}
that this value is exact for random incidence matrices, but that loops have
nevertheless some influence on the localization properties.

So the percolation transition has a drastic effect on the topology of the
random graph and the localization transition changes the behavior of
eigenvectors of its incidence matrix, but as we shall see later, the
transitions have no obvious impact on the moments of the spectral
distribution of a random matrix.  In fact, the situation may look
paradoxical: many relevant quantities for the spectrum of random incidence
matrices (for instance the moments) can be computed by looking at local
structures on the random graph.  For fixed $\alpha$ and large $N$, such
structures are trees with probability 1.  Hence loops that appear for
$\alpha>1$ seem to play no role.  However, the presence of loops is crucial
to ensure that the statistics of finite structures varies smoothly with
$\alpha$.  For example, if, instead of random graphs, one considers random
forests (union of trees)~\cite{bauer00b}, one finds that for $\alpha<1$,
the thermodynamical properties are exactly equal to the ones for the random
graph model, but the transition at $\alpha=1$ (when an infinite tree
appears) changes the distribution of local structures, and for instance the
moments of the spectral distribution are not analytic at $\alpha=1$ for
random forests.

\section{Computation of moments}
%=================================

In this section, we derive Eq.~(\ref{eq:trk=normkplet}), valid for any $N$
and edge parameter $p$, which allows, for a given $k$, to compute directly
$\overline{\Tr \; M^k}$, (i.e. $N$ times the $k$th moment of the density of
states), when $M$ is a random incidence matrix in $\Omega_N$.  A sum rule
for $p=1$ is given.  Then the algorithm is adapted for the variant of the
model with random signs.  Finally we give a compact formula,
Eq.~(\ref{eq:genfuncmom}), for the generating function of moments.

\subsection{Direct computation}
\label{sec:dc}

For any random incidence matrix $M$ in $\Omega_N$, $\Tr\;M^0=N$ and
$\Tr\;M^1=0$, so we may assume that $k \geq 2$.  By definition, $\Tr \; M^k
= \sum_{i_1,\cdots,i_k=1}^N M_{i_1i_2}M_{i_2i_3}\cdots M_{i_ki_1}$. Because
the diagonal matrix elements of $M$ vanish, we can restrict the above sum
and consider only $k$-plet $(i_1,i_2,\cdots ,i_k)$ such that $i_1 \neq i_2,
i_2 \neq i_3,\cdots , i_{k-1}\neq i_k,i_k \neq i_1$. We call such $k$-plets
{\em admissible}.

So, start with an admissible $k$-plet $I=(i_1,i_2,\cdots ,i_k)$. To compute
$\overline{M_{i_1i_2}M_{i_2i_3}\cdots M_{i_ki_1}}$ we argue as follows :
the product $M_{i_1i_2}M_{i_2i_3}\cdots M_{i_ki_1}$ can take only two
values, $0$ or $1$. It has value $1$ if and only if each factor has value
$1$, that is if and only if $\{i_1,i_2\},\{i_2,i_3\},\cdots,\{i_k,i_1\}$
are edges of the graph with incidence matrix $M$. From our definition of
probabilities on the space of incidence matrices, this happens with
probability $p^l$, where $l$ is the number of distinct pairs among
$\{i_1,i_2\},\{i_2,i_3\},\cdots,\{i_k,i_1\}$. Obviously, $l$ depends on
$I$.

If we could find an efficient way to count the number of admissible
$k$-plets $I$ with given $l$, the problem would be solved. We have not been
able to do so.  However, there is a simple way, which we now expose, to
group together families of admissible $k$-plets that are guaranteed to have
the same $l$.

Fix a $k$-plet $W=(v_1,\cdots,v_k)$ of elements of $\{1,\cdots,k\}$ with
the following properties : $i)$ $v_1 \neq v_2,v_2 \neq v_3,\cdots ,v_{k-1}
\neq v_k,v_k \neq v_1$, and $ii)$ if $v_{\beta}>1$, there is a $\beta ' <
{\beta}$ such that $v_{\beta '}=v_{\beta}-1$.  Such a $k$-plet will be
called a {\em normalized} $k$-plet in the sequel.

The first condition is almost the definition
of an admissible $k$-plet, the only difference being that the
members are in $\{1,\cdots,k\}$, not $\{1,\cdots,N\}$. The second
condition says that the order of appearance of elements of
$\{1,\cdots,k\}$ in the sequence $(v_1,\cdots,v_k)$ is the natural
order. Because of these two conditions, $v_1=1$ and $v_2=2$ in any
normalized $k$-plet, but $v_3$ could be 1 or 3. 

By condition $ii)$, the integers appearing in $W$ build a set of the form
$V=\{1,\cdots,n\}$ for a certain $n \leq k$. Let $E$ be the set whose
elements are the (distinct) pairs among
$\{v_1,v_2\},\{v_2,v_3\},\cdots,\{v_k,v_1\}$.

Now choose an injective map $\sigma$ from $V$ to $\{1,\cdots,N\}$. The
number of such maps is $N^{\underline{n}}\equiv N(N-1)\cdots(N-n+1)$. Then
the sequence $I=(\sigma (v_1),\cdots,\sigma (v_k))$ is an admissible
$k$-plet by the injectivity of $\sigma$ and property $i)$ of the sequence
$W$. Moreover, for the same reasons, the number $l$ of distinct pairs among
the $k$ pairs $\{\sigma (v_1),\sigma (v_2)\}, \{\sigma (v_2),\sigma
(v_3)\},\cdots,\{\sigma (v_k),\sigma (v_1)\}$ is exactly $|E|$, the number
of elements of $E$. This number depends on $W$, but not on $\sigma$.

More precisely, there is a one to one correspondence between admissible
$k$-plets $I$ and pairs $(W,\sigma)$.  Note that the source of $\sigma$
depends on $W$.

The bijection involves a simple but useful general algorithm, which we call
the ``label and substitute algorithm''. We use it several times in the
sequel. If $(O,O',O'',\cdots)$ is any finite or infinite list of items
(some items can be repeated), one can label the items in order of first
appearance. This means that the first item receives label $1$, then the
next item different from the first one receives label $2$ and so on. This
gives a one to one map, the ``labeling map''. After that, by replacing each
item in the list by its label, one obtains a sequence of integers, the
``substitution sequence''. It has the property that the first term is $1$
and that if integer $i+1\geq 2$ appears as a term, then integer $i$ has
appeared before. Note that this new sequence is invariant if we apply the
``label and substitute algorithm'' to it.  Note also that the knowledge of
the ``labeling map'' and the ``substitution sequence'' allows to
reconstruct the original sequence.  Let us give an example. The list (eat,
work, eat, sleep, work, eat, work, sleep) leads to the ``labeling map''(eat
$\rightarrow 1$,work $\rightarrow 2$,sleep $\rightarrow 3$), and to the
``substitution sequence'' $(1,2,1,3,2,1,2,3)$.

If $I=(i_1,i_2,\cdots,i_k)$ is an admissible $k$-plet, we apply the ``label
and substitute algorithm''. Then $\sigma$ is the inverse of the ``labeling
map'' of $I$ and $W$ is the ``substitution sequence'' of $I$. The
properties of $I$ making it an admissible $k$-plet and the definition of
the ``label and substitute algorithm'' ensure that $W$ is a normalized
$k$-plet.

Written symbolically, this means that
\[
  \sum_{I}=\sum_{W} \sum_{\sigma}
\] 
where the sum over $I$ is over admissible $k$-plets, the sum over $W$ is
over normalized $k$-plets, and the sum over $\sigma$ is over injective maps
as described above. Inserting $\overline{M_{i_1i_2}M_{i_2i_3}\cdots
M_{i_ki_1}}$ on both sides of this identity, we obtain our first important
formula:
\begin{equation} 
  \label{eq:trk=normkplet} 
  \overline{\Tr \; M^k} = \sum_W N^{\underline{|V|}} p^{|E|}, 
\end{equation} 
where $V$ and $E$ are functions of $W$ as defined above. In this formula,
the $N$ and $p$-dependence are completely explicit. For finite $k$ and
large $N$ this is clearly useful because the $W$'s are defined
independently of $N$ and $p$.

It is not difficult in principle to enumerate normalized $k$-plets in
standard lexicographic order, hence in particular in order of increasing
$|V|$, and then compute for each normalized $k$-plet the value of $|E|$.

We know that 
\[\overline{\Tr \; M^0}=N\] 
and 
\[\overline{\Tr \; M^1}=0.\] 
For $k=2$, the only sequence is $(1,2)$, so 
\[\overline{\Tr \; M^2}=pN^{\underline{2}}.\] 
For $k=3$, the only sequence is $(1,2,3)$, so 
\[\overline{\Tr \; M^3}=p^3 N^{\underline{3}}.\] 
For $k=4$, the sequences are $(1,2,1,2)$, $(1,2,1,3)$, $(1,2,3,2)$ and 
$(1,2,3,4)$, so 
\[
  \overline{\Tr \; M^4}= p^4N^{\underline{4}}+ 2p^2N^{\underline{3}}+
  pN^{\underline{2}}.
\] 
For $k=5$, the sequences are (1, 2, 1, 2, 3), (1, 2, 1, 3, 2), (1, 2, 1, 3,
4), (1, 2, 3, 1, 2), (1, 2, 3, 1, 3), (1, 2, 3, 1, 4), (1, 2, 3, 2, 3), (1,
2, 3, 2, 4), (1, 2, 3, 4, 2), (1, 2, 3, 4, 3), and (1, 2, 3, 4, 5), so
\[
  \overline{\Tr \; M^5}=
  p^5N^{\underline{5}}+5p^4N^{\underline{4}}+5p^3N^{\underline{3}}.
\]   
For $k=6$, one finds 41 sequences leading to
\[
        \overline{\Tr \; M^6}=
          p^6 N^{\underline{6}} + 
        (3 p^6 + 6 p^5) N^{\underline{5}} + 
        (9 p^5 + 6 p^4 + 5 p^3) N^{\underline{4}} + 
        (4 p^3 + 6 p^2) N^{\underline{3}} + 
        p N^{\underline{2}}.
\]   
For $k \geq 7$, by counting the sequences with $|V| \geq k-2$, 
\[
  \begin{array}{l}
    \overline{ \Tr \; M^k} = p^k N^{\underline{k}} +
    \left\{ \left( \frac{k^{\underline{2}}}{2} - 2k \right) p^k + k p^{k-1}
    \right\} N^{\underline{k-1}} \\
    + \left\{ \left( \frac{k^{\underline{4}}}{8} - 
                    \frac{5 k^{\underline{3}}}{6} + k^{\underline{2}} + 5k
             \right) p^k +
             \left( \frac{k^{\underline{3}}}{2} - k^{\underline{2}} - 6k
             \right) p^{k-1} +
             \left( \frac{k^{\underline{2}}}{2} + k \right) p^{k-2}
     \right\} N^{\underline{k-2}}  \\
    + O(N^{k-3})  \\
    = p^k N^k + ( -2k p^k + k p^{k-1} ) N^{k-1} \\
    + \left\{ (k^2 + 4k) p^k - (k^2 + 5k) p^{k-1}
              + \frac{1}{2} (k^2 + k) p^{k-2} \right\} N^{k-2} + O(N^{k-3})
  \end{array}.
\]
With ten days of computation on a workstation, we have obtained the number
of normalized $k$-plets with given $|V|$ and $|E|$ up to $k=18$.  The
results are available upon request to the authors.

\subsection{Sum rule for $p=1$}

We have checked our enumeration against a simple sum rule. We 
put $p=1$ and sum over $|E|$. In this case, with probability $1$,
the random graph becomes complete and
the matrix $M$ is equal to $J-Id$, 
where $Id$ is the $N \times N$ identity matrix 
and $J$ is the $N \times N$ matrix with all entries equal to $1$. But
$J-Id$ has only two eigenvalues, $N-1$ with multiplicity
$1$ and $-1$ with multiplicity $N-1$. So, for $p=1$, 
\[  \overline{\Tr \; M^k} = \Tr \; (J-Id)^k = (N-1)^k+(N-1)(-1)^k.  \] 
The general formula reduces to
\[
  (N-1)^k+(N-1)(-1)^k = \sum_W N^{\underline{|V|}} = \sum_{n}
  N^{\underline{n}}{\CD}_{k,n}
\]
where ${\CD}_{k,n}$ is the number of
normalized $k$-plets $W$ with $|V|=n$.
Going to generating functions, the left-hand side gives 
\[\sum_{k,N}\left((N-1)^k+(N-1)(-1)^k\right)\frac{x^N}{N!}\frac{t^k}{k!}=
e^{-t}(e^{xe^t}+e^x(x-1)),    \]
while the right-hand side gives
\[\sum_{k,N}\frac{x^N}{N!}\frac{t^k}{k!}\sum_{n} N^{\underline{n}}
{\CD}_{k,n} =e^x \sum_{k,n}{\CD}_{k,n}x^n\frac{t^k}{k!}.\]
Hence 
\[
  \sum_{k,n}{\CD}_{k,n}x^n\frac{t^k}{k!}=e^{-t}(e^{x(e^t-1)}+x-1).
\]
By using any symbolic computation software, the computation of all the
${\CD}_{k,n}$'s up to, say, $k=50$ takes only a few seconds.  We can
express ${\CD}_{k,n}$ in terms of standard Stirling numbers of the second
kind, ${\cal S}_{k,n}$.  We shall meet them again in
Sec.~\ref{sec:stirling}.  They can be characterized by the relation $x^k =
\sum_n {\cal S}_{k,n} x^{\underline{n}}$. Using the trick $\overline{\Tr \;
M^{k+1}}+\overline{\Tr \; M^k} = N (N-1)^k = \sum_n {\cal S}_{k,n}
N^{\underline{n+1}}$, one finds ${\CD}_{k+1,n} + {\CD}_{k,n} = {\cal
S}_{k,n-1}$, which gives for $k\geq 1$
\[
   {\CD}_{k,n} = \sum_{r=1}^{k-1} (-1)^{k-1-r} {\cal S}_{r,n-1}.
\]

For $x=1$, we get the generating function of ${\CD}_k = \sum_n
{\CD}_{k,n}$, the total number of normalized $k$-plets.  That is
$e^{(e^t-1-t)}$ and we recognize that the ${\CD}_k$ are so-called
generalized Bell numbers~\cite{sloane}.  To give an idea of the size of the
computations of the first moments, the values of ${\CD}_k$ for $k = 0,
\cdots,18$ are
%%
%% 1,0,1,1,4,11,41,162,715,3425,17722,98253,580317,3633280,24011157,166888165,
%% 1216070380,9264071767,73600798037
%%
1, 0, 1, 1, 4, 11, 41, 162, 715, $3\,425$, $17\,722$, $98\,253$,
$580\,317$, $3\,633\,280$, $24\,011\,157$, $166\,888\,165$,
$1\,216\,070\,380$, $9\,264\,071\,767$ and $73\,600\,798\,037$.  With our
combinatorial interpretation of ${\CD}_k$, it is clear that ${\CD}_k \leq
k!$.  On the other hand, the saddle point evaluation shows that $\log
{\CD}_k \propto k(\log k +o(\log k))$, confirming that the growth of the
computation is extremely rapid.

\subsection{Model with random signs}

The above calculations could be adapted to the variant of the model (see
Sec.~\ref{sec:variant}) defined by a parameter $a\in [0,1]$ where the
non-zero elements of the random matrix are $+1$ with probability $a$ and
$-1$ with probability $1-a$.  Equivalently the edges of the random graph
are dressed with a random sign $\pm 1$.

For an admissible $k$-plet $I=(i_1,i_2,\cdots ,i_k)$, we find
$\overline{M_{i_1i_2}M_{i_2i_3}\cdots M_{i_ki_1}} = p^{l_e} (pb)^{l_o} =
p^l b^{l_o}$ where $b=2a-1$ is the sign asymmetry, $l = l_e+l_o$ is the
number of distinct pairs among $\{i_1,i_2\},\{i_2,i_3\},\cdots,\{i_k,i_1\}$
and $l_e$ (resp. $l_o$) the number of those which are repeated an even
(resp. odd) number of times.  So, Eq.~(\ref{eq:trk=normkplet}) becomes
\begin{equation}
  \overline{\Tr \; M^k} = \sum_W N^{\underline{|V|}} p^l b^{l_o},
  \label{eq:trmks}
\end{equation}
and the first moments can be exactly computed by enumerating all the
normalized $k$-plets:
\begin{eqnarray*}
  \overline{\Tr \; M^2} & = & pN^{\underline{2}}, \\
  \overline{\Tr \; M^3} & = & p^3 b^3 N^{\underline{3}}, \\
  \overline{\Tr \; M^4} & = & p^4 b^4 N^{\underline{4}}
    + 2p^2N^{\underline{3}}+ pN^{\underline{2}}, \\
  \overline{\Tr \; M^5} & = & p^5 b^5 N^{\underline{5}}
    + 5 p^4 b^3 N^{\underline{4}} + 5 p^3 b^3 N^{\underline{3}}, \\
  \overline{\Tr \; M^6} & = &
          p^6 b^6 N^{\underline{6}} + 
        (3 p^6 b^6 + 6 p^5 b^4) N^{\underline{5}} + \\ & &
        (9 p^5 b^4 + 6 p^4 b^4 + 5 p^3) N^{\underline{4}} +  
        (4 p^3 + 6 p^2) N^{\underline{3}} + 
        p N^{\underline{2}}.
\end{eqnarray*}
Of course, the case $a=1$ gives previous results.  If we concentrate on the
even model with $a=1/2$ (for which $b=0$), we see that the summation over
random signs keeps only the walks for which all the edges are visited an
{\em even} number of times.  Consequently for all the odd moments,
$\overline{\Tr \; M^{2k+1}} = 0$ and the density of states is a symmetric
distribution.

\subsection{Generating function}

While very convenient for explicit enumeration, the formula in
Eq.~(\ref{eq:trk=normkplet}) is not always convenient for theoretical
arguments. So we reformulate it.

Starting from a normalized $k$-plet $W$, we have defined two sets $V$ and
$E$. Recall that $V$ is of the form $\{1,\cdots,n\}$ for some $n$ and that
$E$ is made of pairs of distinct elements of $V$. These are exactly the
data for a labeled graph with vertex set $V$ and edge set $E$. In this
framework, $W$ can be interpreted as a closed walk on the graph visiting
all edges (this implies in particular that the graph is connected), the
order of first visit to a vertex respecting the natural order : $W$ starts
at vertex $1$, and its first visit to vertex $2$ occurs before its first
visit to vertex $3$ and so on. Note that in this formulation, many labeled
graphs do not appear (for example, the labeled graphs for which vertices
$1$ and $2$ are not linked by an edge). To resume, only some labeled graphs
and some walks visiting all edges appear. On the other hand, take an
unlabeled connected graph $G$ isomorphic to the one defined by the
normalized sequence $W$. Clearly, it is possible to label it in such a way
that $W$ describes a closed walk on $G$ visiting all edges. This labeling
can be achieved exactly in $| Aut(G)|$ (the order of the automorphism group
of $G$, see Sec.~\ref{sec:aut}) ways. Indeed, two distinct labelings have
to describe a non-trivial automorphism of $G$ because $W$ determines
completely a labeled graph isomorphic to $G$.  Written symbolically, this
means that
\[
  \sum_{W}=\sum_{G} \frac{1}{|Aut(G)|} \sum_{W(G)}
\]
where on the left-hand side the summation is over normalized
$k$-plets whereas on the right-hand side the summation on $G$ is over
isomorphism classes of connected graphs and the summation on $W(G)$ is
over walks 
on $G$ of length $k$ visiting all edges of $G$. So, the fundamental
identity can be rewritten as    
\begin{equation}
  \overline{\Tr \; M^k} = \sum_G \frac{1}{|Aut(G)|}
  N^{\underline{|V(G)|}} \ p^{|E(G)|} W_k(G), 
  \label{eq:trmkg}
\end{equation}
where the sum over $G$ is over isomorphism classes of connected graphs,
$Aut(G)$ is the automorphism group of $G$, $V(G)$ and $E(G)$ are
respectively the vertex set and the edge set of $G$ and $W_k(G)$ is the
number of closed walks of $k$ steps on $G$ visiting all edges of $G$. Note
that graphs with $|V(G)|>N$ or $|E(G)|>k$ do not contribute.  Although
Eq.~(\ref{eq:trmkg}) was established for $k \geq 2$, it is valid for $k
\geq 0$. For $k=1$, as a walk with only one step cannot be closed,
$W_1(G)=0$ and $\overline{\Tr \; M} = 0$.  For $k=0$, as a walk with zero
steps is closed and visits one vertex and zero edges, $W_0(G)=0$ for every
graph except for $\bullet$, the graph with one vertex.  For this graph,
$W_0(\bullet)=1$, leading to $\overline{\Tr \; M^0} = N$.  By convention,
the empty graph is not counted as connected, so it does not appear in
Eq.~(\ref{eq:trmkg}).

The above formula can be used to build a generating function by summing
over $k$. On the left-hand side $\sum_{k\geq 0} \lambda^k
\overline{\Tr \; M^k}=\overline{\Tr \; (1-\lambda M)^{-1}}$ is a 
rational function of $\lambda$ because it is the average of a finite number
(namely $2^{N(N-1)/2}$) of rational functions. To deal with the right-hand
side, define $W_G(\lambda)=\sum_k \lambda ^k W_k(G)$. This counts closed
walks on $G$ of arbitrary length visiting all edges of $G$. Were it not for
the last constraint, life would be easy because the generating function for
the closed walks on $G$ of arbitrary length is simply $\Tr \; (1-\lambda
G)^{-1}$ (in this formula and in later algebraic expressions involving
graphs and traces, we use a convenient abuse of notation : $G$ denotes at
the same time the graph and the incidence matrix obtained by labeling it,
any choice of labeling leads to the same traces). To suppress walks that do
not visit all edges, we can use inclusion-exclusion to obtain:
\begin{equation}
  W_G(\lambda)=\Tr \; \frac{1}{1-\lambda G}-\sum_{G^{(1)}}\Tr \;
  \frac{1}{1-\lambda G^{(1)}} \cdots (-)^l\sum_{G^{(l)}}\Tr \;
  \frac{1}{1-\lambda G^{(l)}}\cdots,
  \label{eq:wgl}
\end{equation}
where $\sum_{G^{(l)}}$ is the sum over all subgraphs of $G$ obtained by
deleting $l$ edges, so this formula ends at $l=|E(G)|$. This expresses
$W_G(\lambda)$ as a finite sum of  
rational functions, so $W_G(\lambda)$ is a rational function. To
summarize, we have proved an identity between rational functions (so
that it is possible to assign values to $\lambda$) :
\begin{equation}
   \overline{\Tr \; \frac{1}{1-\lambda M}} = \sum_G
  \frac{1}{|Aut(G)|} N^{\underline{|V(G)|}} \  p^{|E(G)|} W_G(\lambda).
  \label{eq:genfuncmom}
\end{equation}

\section{Fixed edge probability $p$}
%====================================

In the large $N$ limit with $p$ fixed, F\"uredi and
Koml\'os \cite{furedi} (following work by Wigner \cite{wigner58},
Arnold \cite{arnold} and Juh\'asz \cite{juhasz}) have given a detailed
description of the spectrum :
it consists $N-1$ ``small'' eigenvalues which after a rescaling by
a factor $\sqrt{p(1-p)N}$ build up a semicircle distribution of radius $2$ 
and one large eigenvalue (the Perron eigenvalue) whose distribution is
Gaussian with average $pN+(1-2p)$ and finite variance $2p(1-p)$.  The
appearance of this isolated eigenvalue is due to the non-vanishing
average of the matrix elements.  

We have used this theorem to check our numerical simulations
and our algorithm for computing the moments. We have also computed
numerically level spacings. They obey the Gaussian Orthogonal Ensemble
statistics with good accuracy.

\subsection{Monte-Carlo simulations}

\label{sec:mcp}
Monte-Carlo simulations consist in generating a lot of incidence matrices
of random graphs, computing and studying their spectra.

A matrix is obtained with the following procedure.  It is a symmetric
matrix $M$ of size $N \times N$.  Its diagonal elements are set to 0.  Its
non-diagonal elements $M_{ij} = M_{ji}$ (with $i \neq j$) represent the
edges of the graph.  They are randomly and independently chosen: their
values are 1 with probability $p$ and 0 with probability $1-p$, where $p$
is a fixed parameter.

As the matrix is symmetric, it is diagonalizable and all its eigenvalues
are real numbers.  They are computed using the appropriate routine in Nag
library.  This procedure is repeated with several random matrices.  As we
are interested by the asymptotic behavior when $N$ is large, we compare
different sizes of matrix: $N$= 1000, 2000 and 4000, with 40, 20 and 10
matrices respectively.  So, for each case, the statistical study is done
over 40000 random eigenvalues.

When the edge probability $p$ is fixed, in the spectrum of a given matrix,
we must make the distinction between its largest eigenvalue and the $N-1$
others.  Indeed, all the elements are non-negative ($M_{ij} \geq 0$).
Moreover, when $N$ is not too small, the random graph is connected and the
matrix is irreducible.  Then, the Perron-Frobenius
theorem~\cite{gantmacher} assures that the eigenvalue with the largest
modulus (the Perron eigenvalue) is non-degenerate, positive, and that
the elements of 
the corresponding eigenvector are all positive.

Numerical observations show with great accuracy that the average of
this Perron eigenvalue is $pN$, plus a finite correction which depends
on $p$.  Furthermore its variance (i.e the square of the width of
its distribution) is also finite et depends on $p$.
Indeed, for large $N$, the Perron eigenvector is equal
to $(1\dots 1)^T$, plus small fluctuations, and the eigenvalue is about
$pN$.

\begin{figure}
  \centering
  \includegraphics[width=\figwidth]{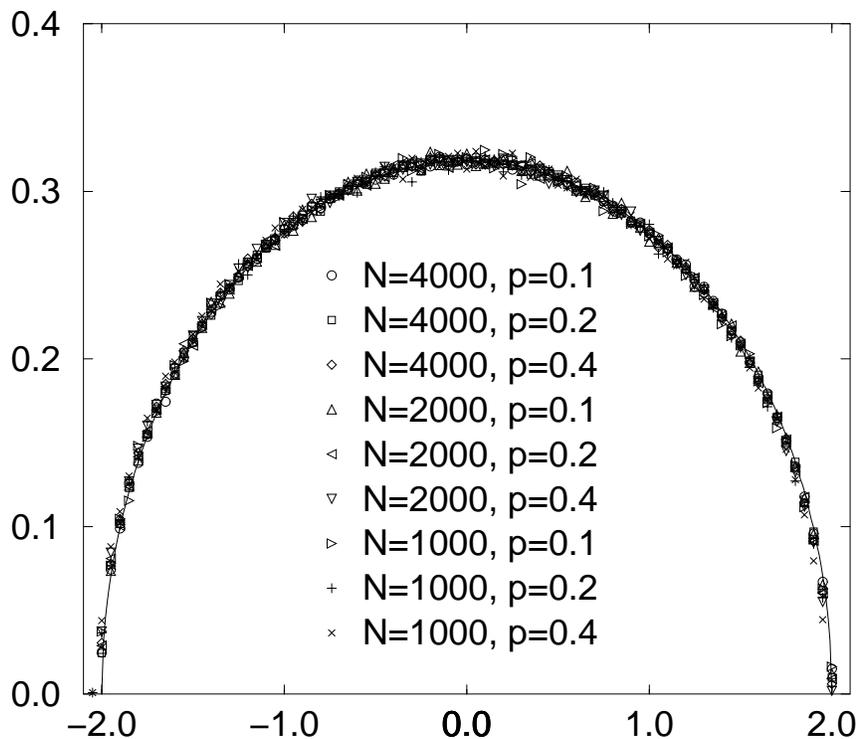}
  \caption{\em
    Histograms of spectra of random graph incidence matrices, for several
    sizes $N$ of matrices and several edge parameters $p$.  The $x$-axis is
    rescaled by $\sqrt{p(1-p)N}$ and the area of each histogram is
    normalized.  All the curves coincide with the semicircle of radius 2,
    drawn with a solid curve.}
  \label{fig:bp}
\end{figure}

As the rest of the spectrum has a large $N$ behavior which is
different we eliminate, in the rest of this
section, the largest eigenvalue of each matrix.
On Fig.~\ref{fig:bp}, histograms of eigenvalues for several
values of $p$ are shown.  To allow comparison, the eigenvalues ($x$-axis)
have been divided by $\sqrt{p(1-p)N}$, and the $y$-coordinates have been
scaled in order to normalize the area of each histogram.  We see clearly
that the asymptotic shape of the rescaled distribution is a semicircle of
radius 2.  In particular, only the variance (or the width) of the
distribution depends on $N$ and $p$, but the shape remains the same.
So already for $N\simeq 1000$ the agreement with the large $N$
estimates in the F\"uredi-Koml\'os theorem is very good. 

\begin{figure}
  \centering
  \includegraphics[width=\figwidth]{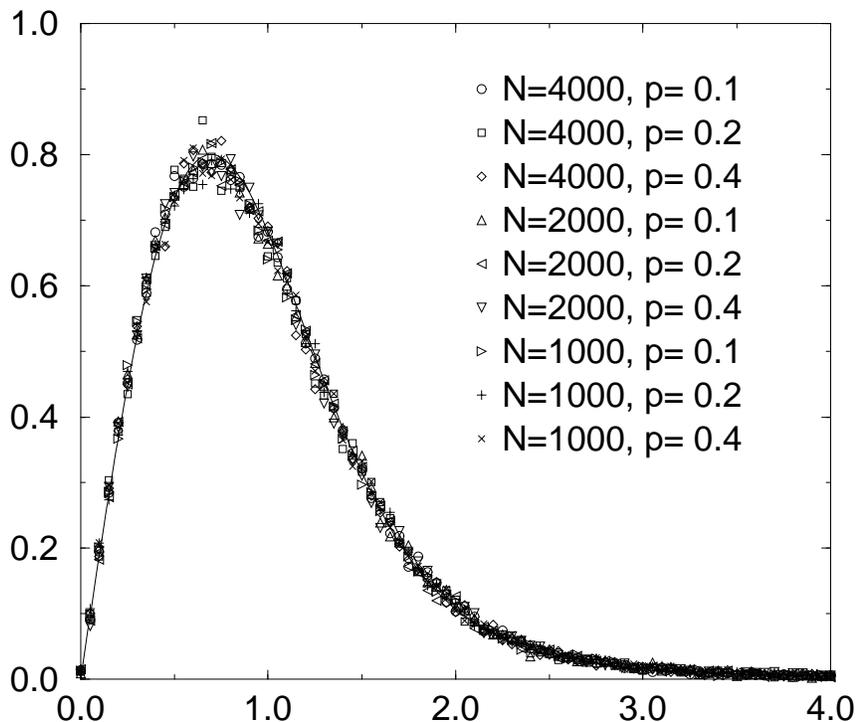}
  \caption{\em
    Histograms of normalized nearest neighbor spacings in spectra of random
    graph incidence matrices, for several sizes $N$ of matrices and several
    edge parameter $p$.  All the curves coincide with the GOE spacing
    distribution distorted by the semicircle law, drawn with a solid curve.}
  \label{fig:sp}
\end{figure}

The nearest neighbor spacing distribution is commonly studied to observe
fluctuations in random spectra.  If $(\lambda_1, \lambda_2, \cdots,
\lambda_N)$ are the eigenvalues in ascending order of a given random
incidence matrix, we define the normalized spacings as
\[
  s_i = \frac{1}{4} \sqrt{\frac{N}{p(1-p)}} (\lambda_{i+1} - \lambda_i),
\]
in order to have $\langle s \rangle = 1$, by omitting the last spacing
$s_{N-1}$ which involves the Perron eigenvalue.  On Fig.~\ref{fig:sp},
histograms of normalized spacings are shown.  We see clearly that they have
the same shape.  This indicates that the large $N$ behavior is independent
of $p$.

To allow the comparison with the spacing distribution of the Gaussian
Orthogonal Ensemble (GOE) of random matrices, we use the ``Wigner surmise''
\[
   q_0(s) = \frac{\pi}{2} s \exp \left( -\frac{\pi}{4} s^2 \right).
\]
It is considered to be an excellent approximation of the GOE
distribution~\cite{mehta}.  However, a direct comparison between the
Monte-Carlo histograms and $q_0(s)$ would give bad results.  Indeed,
$q_0(s)$ describes the distribution of the {\em locally} normalized
spacings $N\rho(\lambda_i) (\lambda_{i+1} -
\lambda_i)$, where $\rho(\lambda)$ is the eigenvalue probability
distribution.   So the probability distribution of $s_i = N (\lambda_{i+1} -
\lambda_i)$ is
\[ 
  q(s) = \int d\lambda \ \rho(\lambda)^2 \ q_0(\rho(\lambda)s).
\]
By taking the semicircle distribution  $\rho(\lambda) = 4/\pi
\sqrt{1-4\lambda^2}$ with a diameter 1 in order to have $\langle s \rangle
=1$, 
\begin{eqnarray*}
q(s) & = & \frac{6}{\pi} \ s \ \exp\left(\frac {-4}{\pi}s^2\right) \ 
           F(\frac{1}{2}, 3, \frac {4}{\pi}s^2) \nonumber \\
     & = & \frac{12}{\pi}\ s \ 
             \exp\left(\frac {-4}{\pi}s^2\right)\ 
             \sum_{k=0}^\infty \frac{\Gamma(k+1/2)}{\Gamma(1/2)}
             \frac{1}{ k! (k+2)!} 
             \left(\frac {4}{\pi}s^2\right)^k,
\end{eqnarray*}
where $F(1/2,3,z)$ is a generalized hypergeometric function.  On
Fig.~\ref{fig:sp}, we see that the Monte-Carlo simulations coincide with
the function $q(s)$, giving good evidence that the incidence matrices of
random graphs are, as expected, in the universality class of GOE.

\subsection{Perturbative expansion for the Perron eigenvalue}

We retrieve quickly the main features of the distribution of the
Perron (i.e. largest) eigenvalue via a perturbative expansion. By definition,
$\overline{M}=p(J-Id)$ where $Id$ is the $N \times N$ identity matrix and
$J$ is $N \times N$ matrix with all entries equal to $1$, that is, $N$
times the projector on $\ket{\Om}=\frac{1}{\sqrt{N}}\left(1 \cdots 1
\right)^T$. So $\ket{\Om}$ is the Perron eigenvector of $\overline{M}$ with
eigenvalue $p(N-1)$. We define $D=M-\overline{M}$. Simple
manipulations~\cite{wigner35} show that an eigenvalue $\la$ of $M$ whose
eigenspace is one-dimensional and not orthogonal\footnote{ This explain why
perturbation theory in $D$ singles out a particular eigenvalue, the Perron
eigenvalue, the only eigenvalue not orthogonal to $\ket{\Om}$ to order $0$
in $D$.} to $\ket{\Om}$ satisfies
\[ 1 =pN \bra{\Om} \frac{1}{\la +p -D} \ket{\Om}. \] 
The perturbative expansion in $D$ gives for the Perron eigenvalue 
\[ \la = p(N-1)+\bra{\Om} D \ket{\Om}  + \frac{1}{pN}\left(
            \bra{\Om} D^2 \ket{\Om} - \bra{\Om} D \ket{\Om}^2 
     \right)+\cdots. \]
Explicit computation yields 
\begin{eqnarray*}
  \overline{\bra{\Om} D \ket{\Om}} & = & 0 , \\
  \overline{\bra{\Om} D^2 \ket{\Om}} - \overline{\bra{\Om} D \ket{\Om}^2} & = &
  p(1-p)(N-1)(N-2)/N.
\end{eqnarray*}
Hence
\[
  \overline{\la^k} = p^k N^k \left\{ 1 + \frac{k}{pN} (1-2p)
                                     + O\left(\frac{1}{N^2}\right) \right\}.
\]

This is enough to show that the distribution of the Perron eigenvalue has
average $\overline\la = pN + 1 - 2p + O(1/N)$ and {\em finite} width
$\overline{\la^2} -\overline\la^2 = O(1)$.  Let us call $\overline{\tr'\,M^k}$
the $k$th moment of the distribution of {\em other} eigenvalues,
\[
  \overline{\tr'\,M^k} \equiv \frac{1}{N-1}
               \left ( \overline{\Tr\;M^k} - \overline{\la^k} \right).
\]
Comparison of our formulae for $\overline{\Tr\;M^k}$ and $\overline{\la^k}$
leads to
\begin{eqnarray*}
  \overline{\tr'\,M}   &=& -p + 2p/N  + O(1/N^2), \\
  \overline{\tr'\,M^2} &=& p(1-p) N + p(3p-2) + O(1/N), \\
  \overline{\tr'\,M^3} &=& -3p^2(1-p)N + O(1), \\
  \overline{\tr'\,M^4} &=& 2 \left[ p(1-p) \right] ^2 N^2 + O(N), \\
  \overline{\tr'\,M^k} &=& O(N^{k-3}) \mbox{ for } k \geq 5.
\end{eqnarray*}
This is of course consistent with \cite{furedi}. Note however that
for the Laplacian matrix (see Appendix~\ref{sec:laplacian}) the hypotheses
of the theorem are not fulfilled because the
diagonal elements are correlated to the rest of the matrix and have
a variance of order $N$. And indeed, the Laplacian has an entirely
different spectral distribution.

\subsection{Comparison with the model with random signs}

The random sign model defined in Sec.~\ref{sec:variant} (where $M_{ij}
= +1$ with probability $ap$, $-1$ with probability $(1-a)p$ and 0 
with probability $1-p$) is also covered by the results in
\cite{furedi}. If $a \neq 1/2$, the ``small'' eigenvalues build a
semicircle of radius $2$ after rescaling by $\sqrt{p(1-b^2p)N}$
(where, as before, $b=2a-1$), 
and the large eigenvalue has a Gaussian distribution
with average $(N-1)bp+(1-b^2p)/b$ and variance $2p(1-b^2p)$.  

As another check of our formul\ae, we give a short proof of the
semicircle distribution for the symmetric ($a=1/2$) random sign model,
where $M_{ij} = \pm 1$ with probability $p/2$ and 0
with probability $1-p$. To compute
$\overline{\Tr\;M^k}$ from Eq.~(\ref{eq:trmks}), we
must keep in the summation only the $W$'s for
which the edges are repeated an even number of times.  Then for the odd
moments, $\overline{\Tr\;M^{2k+1}} = 0$.  For even moments,
$\overline{\Tr\;M^{2k}}$, the maximal number of distinct edges in $W$ is
$k$.  Moreover in the large $N$ limit with $p$ fixed, we retain the $W$'s
with the maximal number of vertices.  This maximum is $k+1$ and is
obtained with $W$'s associated to rooted planar trees with $k$ edges, as
explained later in Sec.~\ref{sec:stirling}.  Then, for $a=1/2$,
\[
   \frac{1}{N}\overline{\Tr\;M^{2k}} = C_k p^k N^k + O(N^{k-1})
\]
where $C_k$ are Catalan numbers, $C_k = (2k)!/[k!(k+1)!]$.
As the Catalan numbers are the moments of the semicircle law of radius 2,
this shows that the density of states is the semicircle law of radius
$2\sqrt{pN}$.

\section{Fixed average connectivity $\alpha$}
%=============================================

In this section, we present results for the large $N$ limit with
$\alpha=pN$ fixed.  After displaying numerical observations, we prove that
the density of states has an infinity of delta peaks for any $\alpha$.
Then we explain how to compute the $2k$-th moment of the density of states
which is a polynomial in $\alpha$ of degree $k$.  Finally, we give bounds
for the coefficients of these polynomials and use the bounds to show that
the spectrum is determined by the moments.

\subsection{Monte-Carlo simulations and observations}

In Sec.~\ref{sec:variant}, we have defined a variant of the model where the
non-zero elements of the random matrix have a random sign: $+1$ with
probability $a$ and $-1$ with probability $1-a$.  We will prove in
Sec.~\ref{sec:ai} that in the large $N$ limit with $\alpha$ fixed, the
moments of the density of states are independent of $a$.  Then it is
expected that the density of states evaluated by our Monte-Carlo
simulations are very similar for those equivalently
obtained~\cite{evangelou92,evangelou92a} for the even model with $a=1/2$.
However as we will clarify some points, we have repeated these simulations.

Monte-Carlo simulations have been done with the same procedure as in
Sec.~\ref{sec:mcp}.  We have seen that, for a fixed edge probability $p$, the
shape of the distribution of eigenvalues is a semicircle.  But when $p$
becomes of order $1/N$, the distribution gets strongly distorted.  To
compare different sizes $N$, let us fix $\alpha = p N$.  It is the average
number of 1's in a given row (or column) of the random matrix.  For the
graph, it is the averaged connectivity (i.e. the average number of
neighbors of a given vertex).

\begin{figure}
  \centering
  \includegraphics[width=\figwidth]{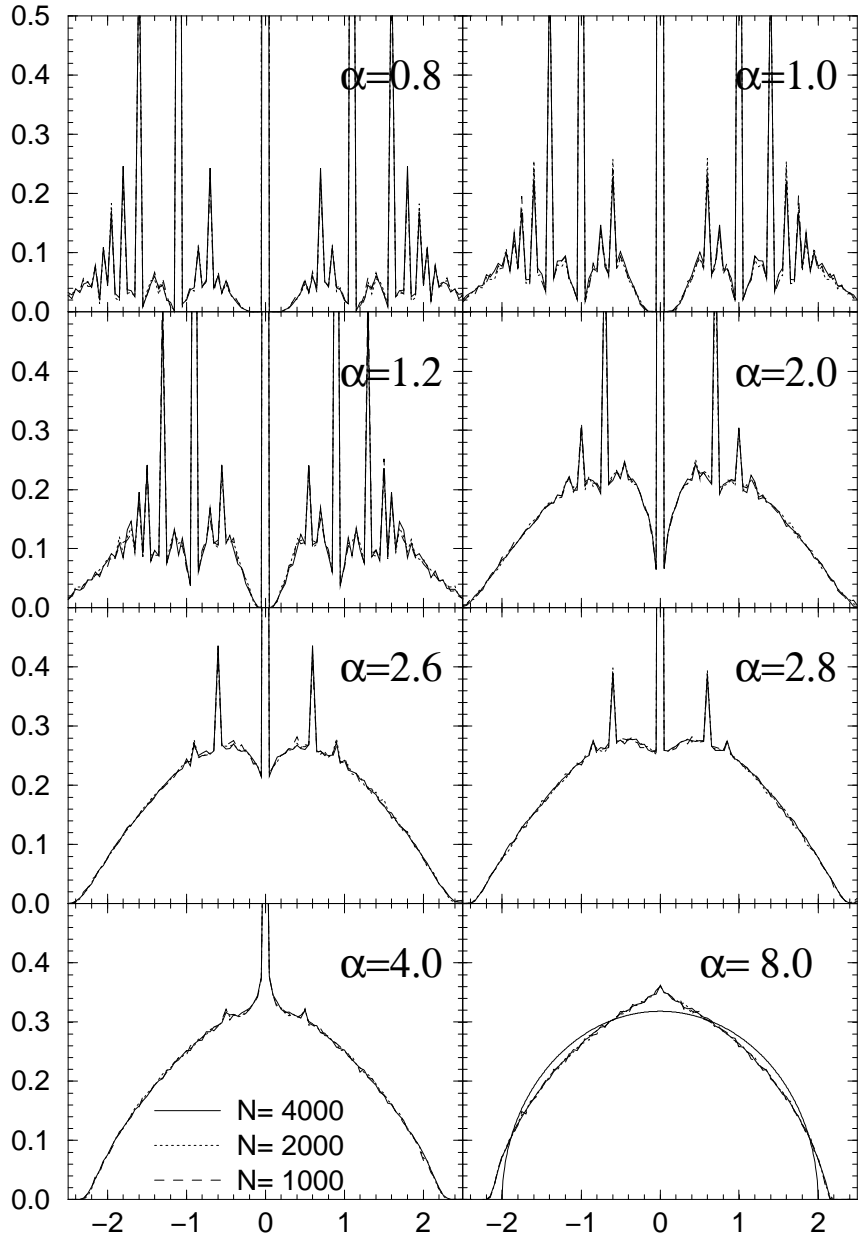}
   \caption{\em
     Histograms of spectra of random graph incidence matrices, for
     different sizes $N$ of matrices and probability parameter
     $p=\alpha/N$, with $\alpha$ fixed.
     The $x$-axis is rescaled by $\sqrt{\alpha}$ and the
     area of each histogram is normalized.  
     For comparison, the semicircle is drawn for $\alpha=8$.
   } \label{fig:ba}
\end{figure}
On Fig.~\ref{fig:ba}, histograms of eigenvalues for several values of
$\alpha$ are shown.  For each value of $\alpha$, three sizes of matrices,
$N=$ 1000, 2000 and 4000, have been simulated with 40, 20 and 10 matrices
respectively.  So, for each case, the statistical study is done over 40000
random eigenvalues.  For fixed $\alpha$, the three curves are superposed,
in the limits of Monte-Carlo fluctuations.  So we can consider that we
observe the asymptotic distribution $d\rho_\alpha$ for large $N$, which
depends only on $\alpha$.  To allow comparison between different
$\alpha$'s, the eigenvalues ($x$-axis) have been divided by
$\sqrt{\alpha}$, and the $y$-coordinates have been scaled in order to
normalize the area of each histogram.

\begin{figure}
  \centering
  \includegraphics[width=\figwidth]{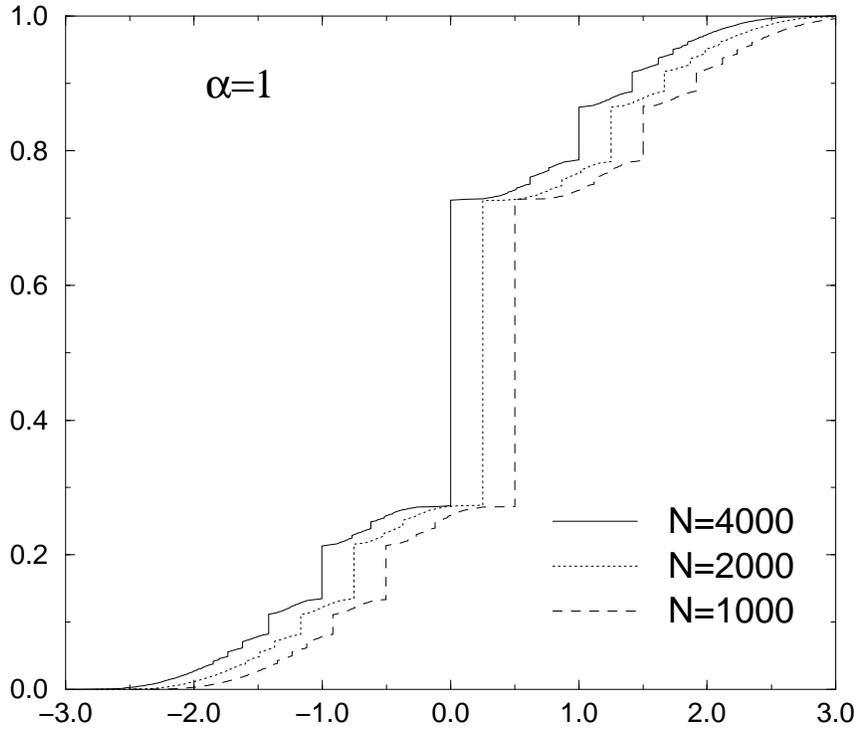}
  \caption{\em
    Cumulative histograms of random graph incidence matrices, for different
    sizes $N$ of matrices and probability parameter $p=\alpha/N$, with
    $\alpha=1$.  Each vertical step represents a delta peak.  As the three
    curves are quite similar, they are shifted along the $x$-axis to be
    more visible.  The largest step is at $x=0$.
  } \label{fig:cumul}
\end{figure}

When $\alpha$ is small, we see a forest of delta peaks.  This has been
previously observed in other models of sparse random
matrices~\cite{kirkpatrick72,evangelou83}.  Their heights are not
representative with this kind of histogram because they depend on the
width of the bin, chosen arbitrarily.  To give a correct representation of
the importance of delta peaks, the cumulative distribution function
(i.e. the integral of the distribution from $-\infty$) is better.  We plot
it for $\alpha =1$ on Fig~\ref{fig:cumul}: each vertical step corresponds
to a delta peak, with a weight equal to the height of the step. As the
heights are comparable between the different sizes, the delta peaks survive
in the limit $N \to \infty$.

We observe that for any $\alpha$ the bigger delta peaks are, in order of
importance, at $x=0$, $\pm 1$, $\pm\sqrt{2}$, $\pm (\sqrt{5} \pm 1)/2$
(golden mean), $\pm\sqrt{3}$, $\pm\sqrt{2 \pm \sqrt{2}}$, etc.  These
values can be recognized as eigenvalues of small trees.

When $\alpha$ increases, the heights of delta peaks decreases, but their
positions along the $x$-axis do not move (before the $\sqrt{\alpha}$
rescaling).  With this kind of histogram, when a delta peak becomes too
small, it seems to disappear because it is drowned in the rest of the
distribution.  But we will show that the spectrum has an infinity of delta
peaks for any $\alpha$.  For large values of $\alpha$, the shape is close
to the semicircle, obtained when $p$ is finite (which corresponds to the
limit $\alpha = \infty$).

In Sec.~\ref{sec:perco}, we have seen that the topology of random graphs
has a percolation transition at $\alpha = 1$.  But, this transition seems
without effects for the distribution of eigenvalues: on Fig.~\ref{fig:ba},
distribution for $\alpha = 0.8$, 1 and $1.2$ are qualitatively similar.  In
particular, we have checked that the height of the delta peak at $x=0$ is
regular around $\alpha = 1$.  The same remarks apply to the localization
transition, conjectured~\cite{evangelou92,evangelou92a} to be for $\alpha
\approx 1.4$.

In contrast, we observe a change of behavior between $\alpha = 2.6$ and
$2.8$, for the distribution $d\rho_\alpha$ in the vicinity of $x=0$.  For
$\alpha \leq 2.6$ (resp. $\geq 2.8$), $d\rho_\alpha(\lambda)$ decreases
(resp. increases) when $\lambda$ goes to $0^+$.  It is difficult to say if
the limit $d\rho_\alpha(\lambda)$ when $\lambda$ goes to $0^+$ is 0 or not
in the small $\alpha$ phase, and $\infty$ or not in the large $\alpha$
phase.  This transition seems to be related to a transition at $\alpha =
e$, where the second derivative of the height of delta peak at $x=0$ as
function of $\alpha$ is discontinuous~\cite{bauer00a}.  Unfortunately, this
discontinuity is too small to be seen in our Monte-Carlo simulations.

We have also studied the distribution of nearest neighbors spacings.  Our
conclusions are similar to those of Evangelou and
Economou~\cite{evangelou92,evangelou92a}: for small $\alpha$, the
distribution numerically coincides with an exponential, and for large
$\alpha$ with the GOE law.  In the vicinity of the localization transition,
$\alpha \approx 1.4$, the distribution interpolates between these two
forms.  On the other hand, we have not been able to reproduce
results~\cite{evangelou92,evangelou92a} concerning the singularity of the
spectral distribution as $\lambda$ goes to 0.

\subsection{Existence of delta peaks}

In this section, we explain that, for any value
of $\alpha$, the distribution has an infinite number of delta peaks.  More
precisely, we show that they are delta peaks at all eigenvalues of finite
trees.  However their heights are exponentially decreasing functions of
$\alpha$; most of them are hidden in simulations by the statistical noise
and the distribution seems to be quite smooth for large $\alpha$.

The delta peaks at tree eigenvalues have, at least, two
origins~\cite{kirkpatrick72}: the small connected components, which are
trees, and small trees grafted on the giant component (percolation
cluster).

The random incidence matrix is block-diagonal, with one block per each
connected component.  As shown by Eq.~\ref{eq:probtree}, the average number
of times a given tree $T$ appears as a connected component of the
random graph is proportional to $N$.  So, for any eigenvalue of the
incidence matrix of any tree with $n$ vertices,
a contribution of height
$\alpha^{n-1} e^{-n\alpha} / |Aut(T)|$ to the corresponding delta peak
appears.  These eigenvalues are algebric numbers, i.e. solutions of a
polynomial equation with integer coefficients of degree at most
$n$.  The height (but not the position) of the peak depends on $\alpha$. 
The height decreases exponentially with $n$, so only eigenvalues of small
trees appear with repetitions in Monte-Carlo simulations.

Furthermore, for $\alpha > 1$, even the giant component contributes to
delta peaks.  To explain why, we first define grafting.  Let $G=(V,E)$ be a
graph, $G'=(V',E')$ and $G''=(V'',E'')$ be two subgraphs of $G$, and $V_0$
be a subset of $V$.  We say that $G$ is obtained by grafting $G'$ on $G''$
along $V_0$ if $V_0 = V' \cap V''$ and $E$ is the disjoint union of $E'$
and $E''$.

If this is the case, suppose moreover that the incidence matrix of $G'$ has
an eigenvector $\phi'$ (with eigenvalue $\lambda$) whose components on
$V_0$ are 0.  Then the vector $\phi$ obtained by extending $\phi'$ to $V$
by 0 is an eigenvector of $G$ with the same eigenvalue $\lambda$.

Now if $G$ is the giant component of a random graph, if $G'$ has a finite
number of vertices and is connected, then $G'$ is a {\em tree} with
probability 1.  Indeed, it is known~\cite{bollobas} that finite connected
subgraphs with loops are suppressed by powers of $N^{-1}$, even if they
belong to the giant component.  For a given tree $T$, the average number of
times $G$ can be obtained by grafting a subgraph $G'$ isomorphic to $T$ on
a subgraph $G''$ along $n$ points is of order $N \alpha^{|E(T)|}
e^{(n-V(T))\alpha}/|Aut(T)|$ for large $\alpha$.  So we have shown that if
$T$ has an eigenvector (with eigenvalue $\lambda$) vanishing on some
vertices, $\lambda$ appears as a delta peak in the spectrum of the giant
component.

\begin{figure}
  \centering
  \includegraphics{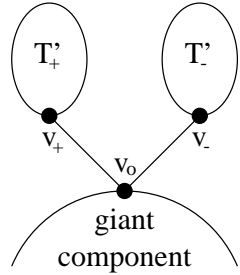}
  \caption{\em
    A symmetric graft on the giant component gives delta peaks in
    the distribution of eigenvalues.}
  \label{fig:ttsym}
\end{figure}

Now if $\lambda$ is an eigenvalue of the incidence matrix of a tree $T'$
with eigenvector $\phi'$, then $\lambda$ also appears as an eigenvalue of
some tree $T$ whose associated eigenvector vanishes on some vertices of
$T$: for instance, choose a vertex $v$ on $T'$, take two copies $T'_+$ and
$T'_-$ of $T'$ and built a tree $T$ by joining $v_+$ and $v_-$ to a new
vertex $v_0$.  Then $\phi = (\phi',0,-\phi')$ on $V(T) = V(T'_+) \cup
\{v_0\} \cup V(T'_-)$ is an eigenvector of $T$ with eigenvalue $\lambda$
and this one can be grafted along $v_0$ (see Fig.~\ref{fig:ttsym}).
So we have shown that the giant component contributes to delta peaks in the
spectrum at all eigenvalues of finite trees.

To summarize, at any eigenvalue of any finite tree, a delta peak appears in
the distribution, with contributions both from small connected components
and from the giant component (for $\alpha > 1$). We do not know if delta
peaks appear at other positions, due to other mechanisms.  On the other
hand, we have not been able to prove that for $\alpha > 1$ the giant
component gives a continuous part in the distribution of eigenvalues.

\subsection{General considerations}
%%%%%%%%%%%%%%%%%%%%%%%%%%%%%%%%%%%
\label{sec:gc}

From an analytical viewpoint, we use equation (\ref{eq:trk=normkplet})
which we rewrite as
\[
  \overline{\Tr \; M^k} =\sum_W N^{\underline{|V|}} N^{-|E|} \alpha ^{|E|}
\]
and recall that on the right-hand side, $E$ and $V$ can be interpreted as
vertices and edges of a connected graph. It is then an elementary
topological fact that $|E|\geq |V|+1$, with equality if and only if the
graph is a tree. For fixed $k$, the possible graphs form a finite set, and
there is no difficulty to take the large $N$ limit :
\[
  \mu_{k}\equiv \lim_{N \rightarrow \infty} \frac{1}{N}\overline{\Tr
  \; M^k} =\sum_{W} \alpha ^{|E|}
\]
where the sum of over normalized $k$-plets $W$ associated to trees. The
existence of the $N \rightarrow \infty$ limit above is a strong indication
of the existence of a limit eigenvalue probability density $d\rho _{\alpha}$,
for which $\mu_{k}$ is just the usual $k^{th}$ moment.

In the same way, in the large $N$ limit with fixed $\alpha$,
Eq.~(\ref{eq:genfuncmom}) becomes
\begin{equation}
   \lim_{N \rightarrow \infty} \frac{1}{N} 
   \overline{\Tr \; \frac{1}{1-\lambda M}} = \sum_T
   \frac{1}{|Aut(T)|} \alpha^{|E(T)|} W_T(\lambda),
   \label{eq:gentree}
\end{equation}
where the sum runs over the trees $T$.  For trees, we observe that the
computation of $W_T(\lambda)$ using Eq.~(\ref{eq:wgl}) can be simplified:
the sum over all subgraphs of $T$ can be reduced~\cite{bauer00} to the sum
over all subtrees obtained by deleting leaves of $T$.

The dominance of trees here and in Sec.~\ref{sec:perco} has a similar
origin.  Though the giant component contains of order $N$ loops, only a
finite number of them are finite : as already noticed before, a finite
connected induced subgraph of a random graph at fixed $\alpha$ is a tree with
probability 1.

On a tree, it takes an even number of steps to make a closed walk, and
$\mu_k=0$ for odd $k$.  This elementary observation implies in fact that
trees can be bicolored, and by standard argument, this shows that trees
have a symmetric spectrum (see Sec.~\ref{sec:sym} for an application of
this property).  As random graphs look locally like trees, it is not too
surprising that they also have a symmetric spectrum (this is of course true
only for $N \to
\infty$).  Later we shall show that the moments $\mu_k$ determine the
distribution.  Then $\mu_k=0$ for odd $k$ implies that the distribution of
eigenvalues is indeed symmetric.

For the even moments, $\mu_{2k}$ is a polynomial of $\alpha$ with degree $k$,
\[
  \mu_{2k} = \sum_l \CI_{k,l} \alpha^l
\]
where $\CI_{k,l}$ is the number of normalized $2k$-plets $W$ associated to
trees with $l$ edges.

\label{sec:ai}
If we consider the variant of the model with random signs (see
Sec.~\ref{sec:variant}), the same arguments apply.  In particular, only
$W$'s associated to trees contribute to the sums.  But in this case, any
edge is visited an even number of times and, following
Eq.~(\ref{eq:trmks}), $\mu_{2k}$ does not depend to $a$.  Then, in the
large $N$ limit with $\alpha$ fixed, the parameter $a$ is irrelevant.

\subsection{Recursion relation for $\CI_{k,l}$}
%%%%%%%%%%%%%%%%%%%%%%%%%%%%%%%%%%%%%%%%%%%%%%%
\label{sec:rr}

\begin{figure}
  \centering
  \includegraphics{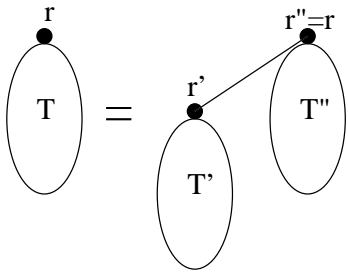}
  \caption{\em Decomposition of $T$}
  \label{fig:tree}
\end{figure}

Consider a normalized $2k$-plet $W$ associated to a tree $T$, with $k \geq
1$.  Recall that $W$ induces a walk covering $T$, hence a labeling of $T$
(vertices are labeled in order of their appearance in the walk). Let $r$ be
the root of $T$, the vertex labeled $1$, where the walk starts. There is
always an edge between vertices labeled $1$ and $2$.  If the edge $\{1,2\}$
is cut, the tree breaks in two trees, $T'$ whose root $r'$ is vertex $2$
and $T''$ whose root $r''$ is vertex $1$ (note that $r''=r$).  The tree $T$
can be seen as trees $T'$ and $T''$ linked by the edge $\{r'',r'\}$, as
shown in Fig.~\ref{fig:tree}.  Note that $T'$ and $T''$ are arbitrary
trees.  They could for instance consist of a single vertex.

The walk $W$ can be decomposed accordingly : the walker starts at vertex
$r''$, walks along edge $\{r'',r'\}$, makes a closed walk on $T'$ (made
possibly of zero steps), walks along edge $\{r',r''\}$, makes a closed walk
on $T''$ (made possibly of zero steps), and so on and so forth, and finally
comes back to vertex $r''$, after covering $T'$ and $T''$.  Let $2k'$
(resp. $2k''$) be the number of steps made on $T'$ (resp. $T''$). We can
glue the small pieces of walks on $T'$ (resp. $T''$) together into a single
walk, covering $T'$ (resp. $T''$) because $W$ covers $T$. This walk is a
sequence of vertices on $T'$ (resp. $T''$), and the ``label and substitute
algorithm'' applied to this sequence leads to a normalized $2k'$-plet $W'$
(resp. to a normalized $2k''$-plet $W''$).

Now, the $2k$-plets $W$ giving rise to the same pair $(W',W'')$ are easily
counted. They differ from each other by the organization of the steps on
edge $\{r',r''\}$. There are $2\overline{m}\equiv 2(k-k'-k'')$ such
steps. If $m'$ (resp. $m''$) is the number of returns of $W'$ (resp. $W''$)
at vertex $r'$ (resp. $r''$), there are exactly
$\bin{m'+\overline{m}-1}{\overline{m}-1}$
(resp. $\bin{m''+\overline{m}-1}{\overline{m}-1}$) possibilities to insert
the $\overline{m}$ steps from $r'$ to $r''$ (resp.  the $\overline{m}$
steps from $r''$ to $r'$). Note that the $-1$ in the above counting comes
from the fact that the last (resp. first) step from $r'$ to $r''$
(resp. from $r''$ to $r'$) is fixed. The total number of visits at vertex
$r''$ is $\overline{m}+m''$, the first term counting visits from the edge
$\{r',r''\}$ and the second visits from $T''$. If $T'$ (resp. $T''$) has
$l'$ (resp. $l''$) edges, $T$ has $l=l'+l''+1$ edges. To summarize, the
number $\CI_{k,l,m}$ of normalized $2k$-plets associated to trees with $l$
edges and containing $m$ times the number $1$ satisfies the following
recursion relation, for $k \geq 1$
\begin{equation}
  \CI_{k,l,m} = 
  \sum
  \CI_{k',l',m'} \ \CI_{k'',l'',m''}
  \bin{m'+\overline{m}-1}{\overline{m}-1} 
  \bin{m''+\overline{m}-1}{\overline{m}-1},
  \label{eq:recurinc}
\end{equation}
where the sums runs over non-negative indices $k'$, $k''$, $l'$, $l''$,
$m'$, $m''$ and $\overline{m}$ with relations $k'+k''+\overline{m}=k$,
$l'+l''=l-1$ and $\overline{m}+m''=m$.

Note that $\CI_{k,l,m}$ vanishes if $l>k$ (every edge is visited at
least twice for a covering closed walk on a tree), if $m>k$ (two
successive terms in a normalized $2k$-plet are distinct, so $1$ cannot 
appear more than $k$ times) and if $m=0$ unless $k=l=0$ (every non void
 normalized $2k$-plet contains $1$). Finally, $\CI_{0,0,0}=1$.
This gives more than enough boundary conditions to compute the
$\CI_{k,l,m}$ recursively.

On a workstation, a day of symbolic computation with this formula gives the
$\CI_{k,l,m}$ as integers up to $k=50$. With 12 digits precision, a Fortran
program goes to $k=120$ in about the same time.  The results are available
upon request to the authors.

\begin{table}
  \centering \tabcolsep 3pt
  \begin{tabular}{c|rrrrr rrrrr}
    $k\backslash l$ & 1 & 2 & 3 & 4 & 5 & 6 & 7 & 8 & 9 & 10 \\ \hline
    1 & 1  \\ 
    2 & 1 & 2  \\ 
    3 & 1 & 6 & 5  \\ 
    4 & 1 & 14 & 28 & 14  \\
    5 & 1 & 30 & 110 & 120 & 42  \\
    6 & 1 & 62 & 375 & 682 & 495 & 132  \\
    7 & 1 & 126 & 1190 & 3248 & 3731 & 2002 & 429  \\
    8 & 1 & 254 & 3628 & 14062 & 23020 & 18928 & 8008 & 1430 \\
    9 & 1 & 510 & 10805 & 57516 & 127029 & 144024 & 91392 & 31824 & 4862 \\
   10 & 1 & 1022 & 31740 & 227030 & 654395 & 968544 & 828495 & 426360 &
        125970 & 16796 
  \end{tabular}
  \caption{\em The number of normalized $2k$-plets associated
           to trees with $l$ edges.}
  \label{tab:ikl}
\end{table}

Note that the index $m$ is not directly relevant for the computation of
moments, because $\CI_{k,l}=\sum_m \CI_{k,l,m}$, but we have not been able
to obtain a closed recursion without the index $m$.  Coefficients
$\CI_{k,l}$ for small $k$ and $l$ are given in Table~\ref{tab:ikl}.

Let us note that each moment is a polynomial in $\alpha$, so that in
particular it does not exhibit any singularity at the percolation or
localization transition.  This is not very surprising, because by the
general algorithm (any $N$ and $p$) $\mu_{k}$ is computed by exploring
connected subgraphs with at most $k$ sites of the random graph. But as
noted before a finite connected induced subgraph of a random graph at fixed
$\alpha$ is a tree with probability 1. To see any trace of a transition,
one would have to look at the behavior of $\mu_{2k}$ for large $k$.  For
instance, we tried to see if complex zeroes of $\mu_{2k}$ in the variable
$\alpha$ (or other related quantities) have a tendency to accumulate near
the real axis.  We have found no conclusive evidence. The rapid growth of
the coefficients $\CI_{k,l}$ makes a numerical study up to $k=50$
difficult, even if we know all numbers exactly.

\subsection{Bounds for $\CI_{k,l}$ }
%====================================

In this section, we will show that $\CI_{k,l}$, the number of normalized
$2k$-plets associated to trees with $l$ edges satisfies the bounds 
\begin{equation}
  \CS_{k,l} \leq \CI_{k,l} \leq C_l \ \CS_{k,l},
  \label{eq:sc}
\end{equation}
where $C_l$ are Catalan numbers defined by
\[
  C_l = \frac {(2l)!} {l! (l+1)!},
\]
and $\CS_{k,l}$ are Stirling numbers of the second kind (i.e. the number of
ways of partitioning a set of $k$ elements into $l$ non-empty subsets)
\label{sec:stirling} which can be computed with the formula
\begin{equation}
  \CS_{k,l} = \frac{1}{l!} \sum_{m=0}^l (-1)^{l-m} \ \bin{l}{m} \ m^k.
  \label{eq:defstirling}
\end{equation}
        
Let $W$ be a normalized $2k$-plet associated to a tree $T$. We view $W$ as
a walk on $T$. Then $W$ allows to put more structure on $T$. First, the
starting point of the walk turns $T$ into a rooted tree. This allows to
talk about sons of a vertex $v$, the root being the initial ancestor. Its
sons are its neighbors, and so on.  Then $W$ also endows the sons of a
vertex with an ordering : the order of first visit. This means that $W$
naturally endows $T$ with a structure of plane rooted tree, with the
convention that the root is at the top of the tree, and the elder son is
always the leftmost one.  Then,
\[
  \CI_{k,l} = \sum_{T_l} \CI_k(T_l)
\]
where the sum runs over the plane rooted trees $T_l$ with $l$ edges, and
$\CI_k(T_l)$ is the number of admissible walks on $T_l$ with $2k$ steps,
where {\em admissible} means starting and finishing at the root of $T_l$
and visiting all the $l+1$ vertices by respecting the order of birth among
brothers (i.e.  a vertex can be visited only if its brothers on its left
have been visited before).

First, we will prove that $\CI_k(T_l^\star) =
\CS_{k,l}$ for a particular tree --- the star-like tree --- which gives
the lower bound of Eq.~(\ref{eq:sc}).  Then, we will prove that
\begin{equation}
  \CI_k(T_l) \leq \CS_{k,l}        \label{eq:ik}
\end{equation}
for any tree $T_l$.  As the number of plane rooted trees with $l$ edges
is $C_l$ (see e.g.~\cite{stanley}), this gives the upper bound of
Eq.~(\ref{eq:sc}).

\begin{figure}
  \centering
  \includegraphics{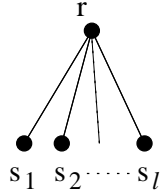}
  \caption {\em Labels of vertices of the ``star'' tree $T_l^\star$.}
  \label{fig:star}
\end{figure}

The star-like tree $T_l^\star$ is made of a root $r$ and $l$ sons:
$\{s_1,s_2,\dots,s_l\}$, see Fig.~\ref{fig:star}.  If $l=0$, then
$\CI_k(T_0^\star) = \delta_{k,0} = \CS_{k,0}$.  So we consider now that $l
\geq 1$.  All the normalized 2k-plet associated to $T_l^\star$ can be
written as $(r, s_{v(1)}, r, s_{v(2)}, \cdots, r, s_{v(k)})$ where $v$ is
an onto map from $[1,k]$ to $[1,l]$, because an admissible walk is made of
$k$ successive double-steps from $r$ to a son and return, and visits all
the sons.  The inverse map $v^{-1}$ makes a partition of $[1,k]$ into $l$
non-empty subsets: $v^{-1}(j) = \{ i | 1 \leq i \leq k ; v(i)=j \} $ for $j
\in [1,l]$.  Conversely, such a partition gives a admissible walk, because
the targets $j$ associated to each subset are uniquely determined by the
birth rule, with the following process: the subset containing 1 is labeled
by $j=1$ and removed, the subset containing the smallest remaining number
is labeled by $j=2$ and removed, etc, up to the last subset labeled by
$j=l$.  This is a one-to-one correspondence between the admissible walks
and the partitions of $[1,k]$ into $l$ non-empty subsets, which are counted
by Stirling numbers of second kind.  Hence, $\CI_k(T_l^\star) = \CS_{k,l}$
for the star-like tree $T_l^\star$.

Let us now consider a given plane rooted tree $T_l$ with $l$ edges. So
Eq.~(\ref{eq:ik}) is true for $k<l$ because $\CI_k(T_l)=0$ (the walk is too
short to visit all the edges) and for $k=l$ because $\CI_l(T_l)=1 =
\CS_{l,l}$ (an admissible walk on $T_l$ of length $2l$ exists and is
completely determined by the birth rule).  Note that this implies that
$\CI_{l,l}=C_l$.  For $k > l$, we will prove Eq.~(\ref{eq:ik}) by
induction, assuming Eq.~(\ref{eq:ik}) to be true for every $(k',l')$ when
$k'<k$.

Let $r$ be the root of $T_l$ and $r'$ the leftmost son of $r$.  As shown on
Fig.~\ref{fig:tree}, we break $T_l$ in three parts, the edge $\{r,r'\}$,
the sub-tree $T'$ with root $r'$ and $l'$ edges and the rest
$T''$, which is a sub-tree with root $r''=r$ and $l''$ edges. We have
$l=l'+l''+1$.  An admissible walk $W$ on $T_l$ with $2k$
steps is composed of $2k'$ steps on $T'$, $2k''$ steps on $T''$
and $2\bar{m}$ steps on the edge $\{r,r'\}$, with
$k=k'+k''+\bar{m}$ and $\bar{m} \geq 1$.

If $T$ is a plane rooted tree, we call $\CI_{k,m}(T)$ the number of
admissible walks on $T$ with $2k$ steps and $m$ returns to its root, and define
\[
  \CH_{k,\bar{m}}(T) \equiv
     \sum_{m=1}^{k}  \CI_{k,m}(T) \bin{m+\bar{m}-1}{\bar{m}-1}.
\]
The arguments used to establish Eq.~(\ref{eq:recurinc}) can be repeated to
show that
\begin{equation}
  \CI_k(T_l) =  \sum_{ \bar{m}, k',k''}^{\bar{m}+k'+k''= k}
                \CH_{k',\bar{m}}(T') \CH_{k'',\bar{m}}(T'').
 \label{eq:iktl}
\end{equation}
Now
\[ 
  \CH_{k',\bar{m}}(T') \leq 
     \sum_{m'=1}^{k'}  \CI_{k',m'}(T') \bin{k'+\bar{m}-1}{\bar{m}-1}
     = \CI_{k'}(T') \bin{k'+\bar{m}-1}{\bar{m}-1},
\]
and by the induction hypothesis, the last term is at most
\[
  \CI_{k'}(T_{l'}^\star) \bin{k'+\bar{m}-1}{\bar{m}-1}.
\]
But this is precisely $\CH_{k',\bar{m}}(T_{l'}^\star)$ because for each
walk on $T_{l'}^\star$, $m'=k'$.  Hence, $\CH_{k',\bar{m}}(T') \leq
\CH_{k',\bar{m}}(T_{l'}^\star)$.  As the same argument holds for $T''$,
$\CI_k(T_l) \leq \CI_k(T_{l',l''}^{\star\star})$ where 
 $T_{l',l''}^{\star\star}$, is the ``bi-star'' tree with
$T'=T_{l'}^\star$ and $T''=T_{l''}^\star$.

It remains to show that $\CI_k(T_{l',l''}^{\star\star}) \leq
\CS_{k,l'+l''+1}$.  If $l'=0$, it is true because the bi-star is simply the
star $T_l^\star$.  As Eq.~(\ref{eq:iktl}) is symmetric by exchange between
$T'$ and $T''$, it is also true for $l''=0$ because
$\CI_k(T_{l',0}^{\star\star}) = \CI_k(T_{0,l'}^{\star\star}) = \CS_{k,l}$.

\begin{figure}
  \centering
  \includegraphics{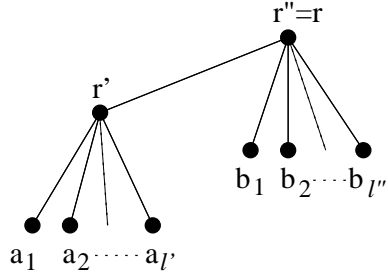}
  \caption {\em Labels of vertices of the ``bi-star'' tree 
            $T_{l',l''}^{\star\star}$.}
  \label{fig:bistar}
\end{figure}

So we have just to consider the case $l'\geq 1$ and $l''\geq 1$.  The
vertices of the bi-star are labeled as shown on Fig.~\ref{fig:bistar}.  The
admissible walks are divided in two classes: (I) the walks finishing by a
step on the right subtree $T_{l''}^\star$ and (II) the walks finishing by a
step from $r'$ to $r''$.

For a walk $W$ of class (I), we call $W^\dagger$ the walk $W$ without the
two last steps; $W^\dagger$ is made of $2k-2$ steps.  The class (I) is
divided in two sub-classes: (I$_1$) $W^\dagger$ has not visited all the
vertices.  Then $W = (W^\dagger, r'',b_{l''})$ and $W^\dagger$ is an
admissible walk on $T_{l',l''-1}^{\star\star}$.  (I$_2$) $W^\dagger$ has
visited all the vertices.  Then $W^\dagger$ is an admissible walk on
$T_{l',l''}^{\star\star}$ and there is $l''$ choices to built $W$ from
$W^\dagger$.  In total, the number of walks of class (I) is
$\CI_{k-1}(T_{l',l''-1}^{\star\star}) + l''
\CI_{k-1}(T_{l',l''}^{\star\star})$, bounded above by $\CS_{k-1,l-1} + l''
\CS_{k-1,l}$.

\begin{figure}
  \centering
  \includegraphics{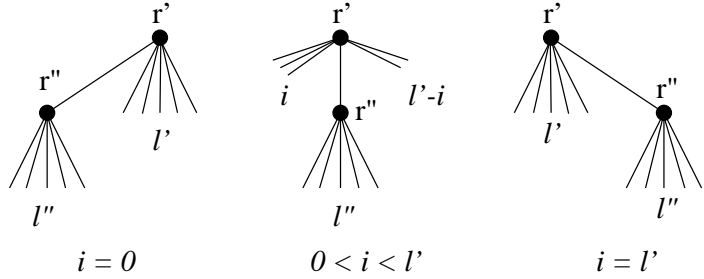}
  \caption{\em The $l'+1$ plane rooted trees corresponding to the class (II).}
  \label{fig:ap1}
\end{figure}

For a walk $W$ of class (II), we call $W^\dagger$ the walk $W$ with the
first and last steps removed.  Then $W^\dagger$ is made of $2k-2$ steps,
starting and finishing at $r'$ (and not $r$) and covering the bi-star.  The
delicate point is the moment of its first visit to $r$.  We have $l'+1$
sub-classes.  In sub-class (II$_0$), the first visit of $r$ is before the
first visit of $a_1$.  In sub-class (II$_i$) with $1\leq i \leq l'-1$, the
first visit of $r$ is after the first visit of $a_i$ and before the the
first visit of $a_{i+1}$.  In sub-class (II$_{l'}$), the first visit of $r$
is after the first visit of $a_{l'}$.  By using the birth rule, each
sub-class corresponds to the admissible walks on one of the plane rooted
trees with $l$ edges drawn on Fig.~\ref{fig:ap1}.  For each of these $l'+1$
trees, the number of steps is $2k-2$, so that the induction hypothesis
applies.  Hence, for class (II), the total number of walks is bounded
above by $(l'+1) \CS_{k-1,l}$.  Then, $\CI_k(T_{l',l''}^{\star\star})
\leq \CS_{k-1,l-1} + (l''+l'+1) \CS_{k-1,l} =
\CS_{k-1,l-1} + l \CS_{k-1,l}$.
But, the Stirling numbers obey to the recursion relation $\CS_{k,l} =
\CS_{k-1,l-1} + l \CS_{k-1,l}$, so that we have reached our goal:
Eq.~(\ref{eq:ik}) is proved and Eq.~(\ref{eq:sc}) follows.

We can now establish two important features of the eigenvalue
distribution $d\rho _{\alpha}$~: it's support is unbounded, but it
Fourier-Laplace transform 
is an entire function, and in particular, the distribution of
eigenvalues is determined by its moments.
 
By using the property of Stirling numbers $n^k = \sum_{l=0}^n \CS_{k,l}
n^{\underline{l}}$ and by summing it with $\sum x^n/n!$, one obtains
\[ 
  \CS_k(x) \equiv \sum_{l=0}^k \CS_{k,l} x^l
         = e^{-x}\sum_{n=0}^{\infty}\frac{n^k}{n!}x ^n.
\]
This leads to a crude bound for the Stirling polynomials when $x$ is
real positive. In fact, $\CS_0(x)=1$, $\CS_{1}(x)=x$, $\CS_{2}(x)=x^2+x$ and
for $k \geq 3$, $\CS_k(x) < k^k+e^{x(k-1)}$ because for $k\geq 3$ and
any $n$, $n^k < k^k+k^n$. 

To see the unboundedness of the support of the eigenvalue distribution, we
observe that the Stirling polynomials $\CS_k(\alpha)$ are the moments of even
order of the even $\al$-dependent probability measure
$\frac{e^{-\al}}{2}\sum_{n=0}^{\infty}\frac{\al ^n}{n!}  (\delta
(x-\sqrt{n})+\delta(x+\sqrt{n}))$ on the real line. The support of this
measure is clearly unbounded. As the moments of $d\rho _{\al}$ are larger
than the Stirling polynomials, the support of $d\rho _{\al}$ has to be
unbounded.

To see the properties of the Fourier-Laplace transform, we just 
show that term by term expansion of
$\int _{-\infty}^{+\infty} d\rho _{\al}(\la  )e^{s\la }$
in powers of $s$ leads to a series with infinite radius of
convergence. For this, we just need to bound 
$ \int _{-\infty}^{+\infty} d\rho _{\al}(\la  ) \frac{\la
^{2k}}{(2k)!} = \mu_{2k}/(2k)! $. We know that this is at most  
$C_{k}\CS_k(\al )/(2k)!= \frac{1}{k!(k+1)!}\CS_k(\al )$. We have seen than 
for $k \geq 3$ this is less than $\frac{k^k+e^{\al(k-1)}}{k!(k+1)!}$
which for large $k$ behaves like $\frac{e^k}{\sqrt{2\pi
k}(k+1)!}=o(R^{-k})$ for any $R$. So, in principle, the knowledge of
the moments determines the probability distribution $d\rho _{\al}$.
However, it is not easy to extract accurate local information on
$d\rho _{\al}$ from the knowledge of a finite number of moments.  
 
\subsection{Special values}

Eq.~(\ref{eq:recurinc}) can be used to compute $\CI_{k,l}$ for special
cases. This is best done using generating functions. Instead of giving the
details, let us just note that this leads to closed forms for $\CI_{k,l}$
for fixed $l$ and any $k$ or for fixed $k-l$ and any $l$, leading to a a
satisfactory description of the borders of the table of moments.  For
$k=0$, $\CI_{0,l} = \delta_{0,l}$.  For $k \geq 1$, the first cases are ~:
\begin{eqnarray*}
  \CI_{k,0} &=& 0, \\
  \CI_{k,1} &=& 1, \\
  \CI_{k,2} &=& 2^k-2, \\
  \CI_{k,3} &=& 3^{k-1}+\om ^k +\bar{\om} ^k -3 \cdot 2^k+2
\end{eqnarray*}
with $\om +\bar{\om}=3$ and $\om \bar{\om}=1$.

On the other border
\begin{eqnarray*}
  \CI_{l,l}   &=& \frac{(2l)!}{l!(l+1)!} \: = C_l, \\
  \CI_{l+1,l} &=& \frac{(2l+2)!}{(l-1)!(l+3)!}, \\
  \CI_{l+2,l} &=& \frac{(2l+4)!}{(l-1)!(l+6)!}\frac{l^2+11l+2}{2},  \\
  \CI_{l+3,l} &=& \frac{(2l+6)!}{(l-1)!(l+9)!}
           \frac{l^4+32l^3+323l^2+232l-48}{6}.
\end{eqnarray*}

That $\CI_{l,l}$ is just the Catalan numbers is not surprising for two
reasons. First combinatorially, for any plane rooted tree with $l$ edges,
$\CI_l(T_l)=1$, so $\CI_{l,l}=C_l$, the number of plane rooted trees.
Second, we know that for fixed $p$, the distribution of eigenvalues is
governed by the semicircle law. It is not surprising that when $\al $ goes
to infinity, the same distribution reappears.  Indeed, the fact that
$\CI_{l,l}=C_l$ is equivalent to the fact that the typical size of
eigenvalues is $\sqrt{\al}$, and that after the rescaling $\la =
\sqrt{\al}x$, $d\rho _{\alpha}(\la)$ converges to
semicircle law
$\frac{1}{2\pi}\sqrt{4-x^2}dx$,
for which the even moments are the Catalan numbers.

The above equations are reminiscent of the meanders problem~\cite{dgg}.  In
particular, we observe that diagonals of Table~\ref{tab:ikl} verify
\[
  \CI_{l+u,l} = \frac{(2l+2u)!} {(l-1)! (l+3u)!} \frac{P_{2u-2}(l)} {u!}
\]
where $P_{2u-2}(l)$ is a polynomial with integer coefficients and with
leading term $l^{2u-2}$.  By conjecturing this form, the coefficients could
be determined for the first values of $u$, from the exact knowledge of the
first $\CI_{l+u,l}$.  Unfortunately, we know no general formula for
$\CI_{k,l}$.

The asymptotics of $\CI_{k,l}$ for fixed $l \geq 2$ and large $k$ are
governed by a simple relation, 
\begin{equation}
  \CI_{k,l} \sim 2 \CS_{k,l} \sim 2 \frac{l^k}{l!}.
  \label{eq:asym}
\end{equation}

As explained in Sec.~\ref{sec:stirling}, $\CI_{k,l} = \sum_{T_l}
\CI_k(T_l)$, and $\CI_k(T_l)$ is maximal for the star tree $T_l^\star$ as
for the bi-star tree $T_{l-1,0}^{\star\star}$ (isomorphic to the star but
with a leaf as root), with $\CI_k(T_l^\star) =
\CI_k(T_{l-1,0}^{\star\star}) = \CS_{k,l}$.  Hence, $\CI_{k,l} \geq 2
\CS_{k,l}$ for $l\geq 2$.  But, for $l$ fixed, we will show that
$\CI_{k,l}$ is asymptotic to this lower bound when $k$ is large, because
among all the $C_l$ plane rooted trees, the contributions of $T_l^\star$
and $T_{l-1,0}^{\star\star}$ become dominant.  Eq.~(\ref{eq:defstirling})
says that $\CS_{k,l} \sim l^k/l!$ for large $k$.

Let $M$ be the incidence matrix of a given plane rooted tree $T_l$ with $l$
edges: the number of admissible walks with $2k$ steps, $\CI_k(T_l)$, is
bounded above by $\Tr\; M^{2k}$ which counts the closed walks on $T$ with
$2k$ steps.  So $T_l$ contributes for large $k$ only if the Perron-Frobenius
eigenvalue of $M$ is not smaller than $\sqrt{l}$.

A tree can be bicolored: if we use black (B) and white (W) as the colors, each
vertex is either black or white, in such a way that edges connect only vertices
with opposite colors.  Then, for an eigenvector $\phi$ with
eigenvalue $\lambda$, the vector $\phi'$, defined by
$\phi'_i = \phi_i$ on black vertices and $-\phi_i$ on white vertices, is
eigenvector associated to eigenvalue $-\lambda$.
Hence, the spectrum is symmetric. \label{sec:sym}

As $T_l$ has $l$ edges, $\Tr M^2 = \sum_{i,j} M_{i,j}^2 = 2l$, and for
the spectrum of $M$, $\sum_i \lambda_i^2 = 2l$.  We see that $T_l$
contribute only if its spectrum is $\{-\sqrt{l}, 0, \sqrt{l} \}$ where 0 is
$l-2$ times degenerated.  By noting $\Phi$ the Perron-Frobenius
eigenvector, associated to $\sqrt{l}$, and $\Phi'$ the
eigenvector associated to $-\sqrt{l}$, in this case
\[
  M_{i,j} = \sqrt{l} \left( \Phi_i \Phi_j - \Phi'_i \Phi'_j \right).
\]
As $\Phi'$ is obtained by inverting the signs of $\Phi$ on white vertices,
$M_{i,j} = 0$ when $i$ and $j$ have the same color and $M_{i,j} = 2\sqrt{l}
\Phi_i \Phi_j$ when $i$ and $j$ have different colors.  The Perron theorem
assures that $\Phi_i > 0$ for all $i$.  Hence, $M_{i,j} = 1$ when $i$ and
$j$ have different colors.  As loops are forbidden in a tree, one of the colors
must color only one vertex.  So $M$ must be the incidence matrix of a
tree isomorphic to the star $T_l^\star$, which prove Eq.~(\ref{eq:asym}).

Using the saddle point approximation, it is not difficult to show that for
large $k$ and $l$ with $k/l$ fixed, $\CS_{k,l}$ grows faster than any
exponential of $l$.  In this regime, $\log \CI_{k,l} \sim \log \CS_{k,l}$
because $\log C_l \sim l \log 4 \ll \log \CS_{k,l}$.  But the asymptotics
of $\CI_{k,l}$ seem much more complicated.

\section{Conclusion}
%===================

In this paper we have made a detailed study of the spectral density
of large random graphs incidence matrices,
both in the fixed edge probability $p$ and in the fixed average 
connectivity $\alpha$ limits. 

For fixed $p$, our results are:
\begin{itemize}
  \item A simple general algorithm to compute arbitrary moments
  (polynomials in $N$ and $p$), easy to implement on a computer (but of
  almost factorial growth), leading to an explicit form for the first 18
  moments.

  \item Numerical diagonalizations of Monte Carlo samples of random
  matrices and analytical arguments to show that when the random signs are
  not chosen with a symmetric probability, the main change is that a
  ``Perron Frobenius'' eigenvalue has to be eliminated to recover the
  semicircle law and the GOE results.
\end{itemize}

For fixed $\alpha$ (the finite connectivity limit), our main contributions
are:
\begin{itemize}
  \item Numerical diagonalizations of Monte Carlo samples of random
  matrices for different values of $\alpha$, with a curious observation for
  $\alpha \simeq 2.7$.

  \item A general proof that the spectrum contains delta peaks at all
  eigenvalues of tree incidence matrices, and that, slightly surprisingly,
  they receive non-vanishing contributions even from the infinite cluster
  when $\alpha > 1$.

  \item A recursion relation of combinatorial origin to compute the
  moments. Odd moments vanish, and we have explicitly computed the first
  120 even moments with the help of a computer. Our algorithm counts
  objects which are related to many other tree enumeration problems of
  independent interest.

  \item A proof that the growth of the moment is slow enough so that they
  determine the spectral distribution entirely.

  \item The moments are insensitive to the random signs, and so is the
  spectrum by the previous remark.

  \item The moments are polynomials in $\alpha$. In particular, they do not
  exhibit any singularity at the classical or quantum percolation
  transitions.

  \item However, we have given a general formula (\ref{eq:gentree}) to
  prepare the ground for a more refined study\cite{bauer00a} of the delta
  peak at the eigenvalue $\lambda=0$ in the spectral distribution. This
  delta peak is directly relevant to quantum percolation and is
  non-analytic in $\alpha$ at $\alpha =e$, which we believe is connected to
  the curious numerical observation at $\alpha \simeq 2.7$ mentioned above.
\end{itemize}

One of the natural continuations of this work would be a careful
numerical and/or analytical study of moments of large order, to see 
if the percolation transitions have a measurable impact on global
characteristics of the spectral density.

From a more mathematical point of view, and because of their close
connection with delta peaks in the spectra of random graphs, it might
be of interest to be able to characterize the eigenvalues of tree
incidence matrices of a given size and how their distribution and
spacings evolve for large sizes.   

\vspace{.5cm}

{\bf Acknowledgments} We thank one of the referees for pointing
reference \cite{furedi} to us. 

\appendix\section{Random Laplacian matrix}
%==========================================

\label{sec:laplacian} 
In this appendix, we give results for the Laplacian matrix on random graphs.
These are obtained using an adaptation of methods previously described.

The random Laplacian matrix $L$ of size $N$ is defined as $L = D - M$ where
$M$ is a random incidence matrix (see Sec.~\ref{sec:model}) with vanishing
diagonal elements and $D$ is the diagonal matrix whose element $D_{ii} =
\sum_{j\neq i} M_{ij}$ is the connectivity of the vertex $i$ in the random
graph associated to $M$.  So all rows or column sums of $L$ vanish.

The Perron-Frobenius theorem applies to $N-L$: the Perron eigenvector is
the uniform eigenvector $(1,\cdots,1)^T$, associated to the eigenvalue
$\lambda_P=0$ for $L$.  It follows that the spectrum of $L$ is real and
non-negative.  Moreover the multiplicity of 0 is the number of
connected components.

To compute $\overline{\Tr\ L^k}$, the method explained in Sec.~\ref{sec:dc}
is adapted to the Laplacian.  After expanding $(D-M)^k$, and eventually
using the invariance of the trace by cyclic permutation, we must compute
several terms individually.  Of course, the diagonal elements $D_{ii}$ need
a special treatment.  With the trick $D_{ii} = \sum_{j\neq i} M_{ij} =
\sum_{j\neq i} M_{ij} M_{ji}$, $D_{ii}$ becomes a double-step $(i,j,i)$ in
the description as a walk on a graph.  Now, an admissible $k$-plet is made
up of single-steps (for $M$) and double-steps (for $D$).  After
enumeration,
\begin{eqnarray*}
  \overline{\Tr\ L  } & = & p N^{\underline 2}, \\
  \overline{\Tr\ L^2} & = & p^2 N^{\underline 3} + 2 p N^{\underline 2}, \\
  \overline{\Tr\ L^3} & = & p^3 N^{\underline 4} + (6p^2-p^3) N^{\underline 3}
      + 4 p N^{\underline 2}, \\
  \overline{\Tr\ L^4} & = & p^4 N^{\underline 5} 
      + (12 p^3 - 3 p^4) N^{\underline 4} + (25 p^2 - 6 p^3) N^{\underline 3}
      + 8 p N^{\underline 2}.
\end{eqnarray*}
A computer program can expand these results for a dozen of $k$.
They can be checked with sum rules: for $p=1$ the matrix is
deterministic and $\overline{\Tr\ L^k} = (N-1) N^k$.

For finite $N$ there are in general several connected components, hence the
Perron eigenvalue $\lambda_P=0$ appears with multiplicity.  However for
fixed $p$ the random graph is connected in the large $N$ limit, and
$\lambda_P=0$ appears only once.  So it is reasonable to consider centered
moments $m_k = \overline{(\lambda - \overline{\lambda})^k}$ where the means
run on the other $N-1$ eigenvalues.  From previous equations,
\begin{eqnarray*}
  m_2 & = & 2 p (1-p) N, \\
  m_3 & = & 4 p (1-p) (1-2p) N, \\
  m_4 & = & p (1-p) [p(1-p)(9N-42)+8] N.
\end{eqnarray*}

In the large $N$ limit with $p$ fixed, $m_3 / m_2^{3/2}$ goes to 0 and
$m_4/m_2^2$ goes to $9/4$.  The limit distribution has a bell-like
shape\cite{biroli99,cavagna99}, intermediate between the semicircle and
the Gauss law for which $m_4/m_2^2$ is 2 and 3 respectively.

In the large $N$ limit with $\alpha = pN$ fixed, the $k$-plets contributing
to the dominant term of order $N$ are associated to trees, as described in
Sec.~\ref{sec:gc}.  So the $k$th moment is a polynomial in $\alpha$ of
degree $k$,
\[
  \mu_k \equiv \lim_{N\to\infty} \frac{1}{N} \overline{\Tr\ L^k} 
        = \sum_l \CL_{k,l} \alpha^l,
\]
where $\CL_{k,l}$ is the number of normalized $k$-plets (with single or
double steps) associated to trees with $l$ edges.  Note that now $k$ is not
constrained to be even, due to the double steps.  Following the
arguments and  notations of Sec.~\ref{sec:rr}, we call
$\CL_{k,l,m}$ the number of normalized $k$-plets associated to trees with
$l$ edges and containing $m$ times the number 1 (i.e. with $m$ return to
the root of the associated tree).

\begin{table}
  \centering \tabcolsep 4pt
  \begin{tabular}{c|rrrrr rrrrr}
    $k\backslash l$  & 1 & 2 & 3 & 4 & 5 & 6 & 7 & 8 & 9 & 10 \\ \hline
 1 & 1  \\
 2 & 2 & 1  \\
 3 & 4 & 6 & 1  \\
 4 & 8 & 25 & 12 & 1  \\
 5 & 16 & 90 & 85 & 20 & 1  \\
 6 & 32 & 301 & 476 & 215 & 30 & 1  \\
 7 & 64 & 966 & 2345 & 1722 & 455 & 42 & 1  \\
 8 & 128 & 3025 & 10696 & 11659 & 4928 & 854 & 56 & 1  \\
 9 & 256 & 9330 & 46453 & 71082 & 43779 & 12012 & 1470 & 72 & 1  \\
 10 & 512 & 28501 & 195340 & 404540 & 342642 & 135357 & 26040 & 2370 & 90 & 1
  \end{tabular}
  \caption{\em The number of normalized $2k$-plets, with single or double
           steps,  associated to trees with $l$ edges.}
  \label{tab:lkl}
\end{table}

In comparison with Sec.~\ref{sec:rr}, as the steps on the first edge
$\{r',r''\}$ are single or double, we call $u'$ (resp. $u''$) the
number of such steps finishing on $r'$ (resp. $r''$), i.e. single steps
$(r'',r')$ and double steps $(r',r'',r')$ (resp. $(r',r'')$ and
$(r'',r',r'')$).  Then for $k\geq 1$, $\CL_{k,l,m}$ satisfies the recursion
relation
\[
  \CL_{k,l,m} = \sum \CL_{k',l',m'} \CL_{k'',l'',m''}
                \bin{m'+u'-1}{u'-1} \bin{m''+u''-1}{u''-1} \bin{u'+u''-1}{u'},
\]
where the sum runs over non-negative indices $k'$, $k''$, $l'$, $l''$,
$m'$, $m''$, $u'$ and $u''$ with relations $k'+k''+u'+u''=k$, $l'+l''=l-1$
and $u''+m''=m$, with the convention that the first binomial coefficient
must be taken as 1 (and not 0) when $m'=u'=0$.  To start the recursion
relation, we need the boundary conditions for $k=0$: $\CL_{0,l,m} =
\delta_l\delta_m$.
 
\begin{table}
  \centering \tabcolsep 4pt
  \begin{tabular}{c|rrrrrr}
    $k\backslash l$  & 1 & 2 & 3 & 4 & 5 & 6 \\ \hline
 2 & 2  \\
 3 & 4  \\
 4 & 8 & 9  \\
 5 & 16 & 50 \\
 6 & 32 & 205 & 56  \\
 7 & 64 & 742 & 574  \\
 8 & 128 & 2513 & 3864 & 431 \\
 9 & 256 & 8178 & 21532 & 6906 \\
10 & 512 & 25941 & 107800 & 68455 & 3942 \\
11 & 1024 & 80894 & 504394 & 540782 & 90508 \\
12 & 2048 & 249337 & 2255128 & 3739054 & 1240360 & 42136 
  \end{tabular}
  \caption{\em  Coefficients $\CL_{k,l}^{(c)}$ of the centered moments of
  the random Laplacian spectral distribution}
  \label{tab:lklc}
\end{table}

By summing on $m$, we retrieve the coefficients $\CL_{k,l} =
\sum_m \CL_{k,l,m}$.  The first of these are given in Table~\ref{tab:lkl}.
As the average $\mu_1=\alpha$ is not zero, we compute the {\em centered}
moments $m_k$, which are still polynomial in $\alpha$ of degree the integer
part of $k/2$.  The coefficients $\CL_{k,l}^{(c)}$ are given in
Table~\ref{tab:lklc}.  The shape of the distribution evolves with $\alpha$:
in particular, it becomes non symmetric as the odd centered moments do not
vanish.  Again, we observe that the results for large $N$ with fixed $p$
are found by keeping the last diagonal of Table~\ref{tab:lkl} and
Table~\ref{tab:lklc}, corresponding to the large $\alpha$ limit.

\end{document}